\newcommand{\la}[1]{\mbox{$
\lefteqn{ \mbox{\,\, \tiny #1}}$} \label{#1}}
\documentclass[article,twocolumn,superscriptaddress,preprintnumbers,showpacs,floatfix,amsmath,amssymb]{revtex4}

\usepackage{graphicx}
\usepackage{dcolumn}
\usepackage{bm}
\usepackage{amsmath}
\usepackage{amssymb}
\usepackage{mathrsfs}
\usepackage{bbm}
\usepackage{epsfig}

\usepackage{wrapfig}
\usepackage{eurosym}
\usepackage[usenames,dvipsnames]{color}
\usepackage[yyyymmdd,hhmmss]{datetime}

\newcommand{\be}{\begin{eqnarray}}
\newcommand{\ee}{\end{eqnarray}}

\renewcommand{\vector}[1]{{\boldsymbol{#1}}}
\newcommand{\eastar}{\end{eqnarray*}}

\newcommand{\ce}{{\mathbbmss e}}

\begin{document}

%
%
%

\title{ ON SPINORS, STRINGS, INTEGRABLE MODELS  \\ AND DECOMPOSED YANG-MILLS THEORY  }

\author{Theodora Ioannidou}
\email{
ti3@auth.gr
}
\affiliation{
Faculty of Civil Engineering,  School of Engineering, 
Aristotle University of Thessaloniki, 54249, Thessaloniki, Greece
}
\author{Ying Jiang}
\email{yjiang@shu.edu.cn}
\affiliation{Department of Physics, Shanghai University, 
Shangda Rd. 99, 200444 Shanghai, P.R. China}
\author{Antti J. Niemi}
\email{Antti.Niemi@physics.uu.se}
\affiliation{Department of Physics and Astronomy, Uppsala University,
P.O. Box 803, S-75108, Uppsala, Sweden}
\affiliation{
Laboratoire de Mathematiques et Physique Theorique
CNRS UMR 6083, F\'ed\'eration Denis Poisson, Universit\'e de Tours,
Parc de Grandmont, F37200, Tours, France}
\affiliation{Department of Physics, Beijing Institute of Technology, Haidian District, Beijing 100081, P. R. China}

\begin{abstract}
\noindent
This paper deals with  various interrelations between strings and surfaces in three dimensional ambient space, two dimensional integrable models and  two  dimensional and four dimensional  decomposed   SU(2) Yang-Mills theories. 
Initially,  a spinor version of the Frenet equation is introduced in order to describe 
the differential geometry of static three dimensional string-like structures. 
Then its relation to the structure of the \underline{su}(2) Lie  algebra valued Maurer-Cartan one-form is presented; while by introducing time evolution of the string  a Lax pair is obtained,  as an integrability condition. 
In addition, it is  show how the Lax pair of the integrable nonlinear Schr\"odinger equation  becomes embedded into the Lax pair of the time extended spinor Frenet equation and it is described how a spinor based projection operator  formalism  can be used to construct the  conserved quantities,  in the case of the nonlinear Schr\"odinger equation. 
Then the Lax pair structure of the time extended spinor Frenet equation is related  to properties of flat connections in a two dimensional  decomposed  SU(2) Yang-Mills  theory. In addition, the connection between the  decomposed Yang-Mills and  the  Gau\ss-Godazzi equation that describes surfaces  in three dimensional ambient space is presented.  
In that context  the relation between isothermic surfaces and  integrable models is discussed.
 Finally,  the utility of the Cartan approach to differential geometry is considered.
In particular, the similarities between the Cartan formalism and the structure of both two  dimensional and four dimensional decomposed SU(2) Yang-Mills theories are discussed,  while the description of   two dimensional integrable models  as embedded structures in the four dimensional  decomposed  SU(2) Yang-Mills theory are presented.
\end{abstract}


\maketitle

\section{Introduction}  
The immersion of a string in the
three dimensional Euclidean space $\mathbb R^3$ is a classic  
subject in differential geometry \cite{frenet}, \cite{spivak}.  
Strings have no intrinsic geometry. They differ from each other 
only in the way  they twist and bend in the ambient space. 
The Frenet equation constructs a string in $\mathbb R^3$
entirely from the knowledge of this extrinsic geometry. 
The solution describes  a dreibein field {\it i.e.} a SO(3) matrix transports 
along the string, in terms of its local curvature and  torsion;
one of the dreibein fields is tangent to the string, while the other 
two constitute a zweibein   on the normal planes of the string. 

The embedding of a two dimensional Riemann surface 
in $\mathbb R^3$ is an equally classic subject. It is governed  
by the Gau\ss-Codazzi equation \cite{spivak}-\cite{pressley}. This equation 
relates closely to the concept of integrability. 
Recall that, in the case of an embedded pseudosphere  \cite{bour},  the integrable 
sine-Gordon equation first appeared as a decomposition of the Gau\ss-Codazzi equation.
Various other models with integrable 
dynamics have been subsequently investigated from this perspective 
\cite{lund}-\cite{rogers}. Of  contemporary interest are different decomposed versions of the
Gau\ss-Codazzi equation that describe isothermic surfaces, and their
relations to known integrable models in two dimensions
\cite{pressley}, \cite{sym95}-\cite{burstall}.
The connection between integrable models and  loop groups should also
be mentioned \cite{uhlenbeck}.

However, most Riemann surfaces are neither pseudospheres nor isothermic manifolds. 
In the general case, the description of a Riemann surface  embedded in $\mathbb R^3$ 
continues to remain beyond the realm of the traditional theory of integrable systems.

Here we are interested in the generic time evolution of a string in three dimensions.
Its time evolution sweeps a Riemann surface that is embedded in $\mathbb R^3$.
Thus the Hamiltonian dynamics should fundamentally relate to the Gau\ss-Codazzi equation.  
In fact, the Frenet equation that describes the string at a fixed time, 
corresponds to an auxiliary linear problem in the 
theory of  integrable models. Moreover, both the nonlinear Schr\"odinger 
equation and the modified KdV equation have  been 
extensively studied, in connection of  the motion of vortex filaments and 
other regular string-like structures in three space dimensions \cite{hasimoto}-\cite{langer1}. 

Even thought not explicitly addressed here, among the motivations of the present work is to  develop a Hamiltonian dynamics that
describes proteins  modeled as discrete piecewise linear polygonal chains. 
A biologically active protein is in a space filling 
collapsed phase {\it i.e.} it's  geometry is fractal. Thus standard 
techniques of embedding theory do not apply. Novel techniques need to be 
developed. The present article is a step towards 
the pertinent formalism, it addresses regular differentiable 
strings with no fractal affiliation, but using formalism that hopefully enables the
description of dynamical fractal string-like objects \cite{danielsson}-\cite{nora}.

Universality arguments allege that only the Polyakov action  \cite{polyakov} should
remain relevant in the high energy limit of a (relativistic) string. 
However, at lower energies  the corrections can not be ignored. 
Such an example is  the extrinsic curvature term \cite{poly}. 
Here we shall be interested in systematic inclusion of such
additional corrections, that describe  the  bending and twisting motions of the string. 
In particular, the relativistic case   and the low energy case  is considered when the dynamics becomes subject to the Galilean invariance \cite{oma1}. 
This is because the corresponding energy functions  have many applications to Physics. Examples include vortex dynamics in  superconductors; fluids and cosmic 
strings \cite{langer2}-\cite{kibble};  polymers, proteins,  and their folding dynamics \cite{danielsson}-\cite{nora}. 

Moreover, beyond Physics,   the Frenet equation has  numerous applications. 
Examples include robotics, computer graphics and virtual reality, 
aeronautics and astronautics \cite{hanson}, \cite{kuipers}.
In these applications the ``gimbal lock", a coordinate  singularity in Euler angles, often appears as a nuisance that needs to be overcome. 
For this,  in lieu of the dreibein Frenet frame description, a quaternionic formulation is 
commonly preferred to describe how  rotation matrices become transported 
along strings and trajectories.

Similarly, the isothermic surfaces and, more generally, isothermal (conformal)
coordinate representations of  embedded Riemann surfaces  are  studied extensively; also, from the point of view of three dimensional visualization \cite{bobenko}.  
There is a wide range of applications such as:
structural mechanics, architecture, industrial design and  so forth \cite{visa}. 
In this setting, a quaternionic representation of rotations is similarly often advantageous  \cite{bobenko2}.

In this paper  a different approach is introduced and develop in order to describe strings, their  time evolution, and the two dimensional Riemann 
surfaces that are  swept by this time evolution.  
Instead of the classic dreibein or the more flexible  quaternionic realization of the Frenet equation,  a string in  $\mathbb R^3$ is represented by 
a two component complex spinor. 
Then the dreibein Frenet equation becomes a two component spinor Frenet
equation describing  the dynamics of a spinor along the string. 
This spinor-based representation of a string has various 
conceptual and technical advantages over both  the conventional dreibein approach  and  its quaternionic modernization. 
Note that despite its simplicity and  apparent advantages, 
a spinor representation of the Frenet equation has been studied  only sparsely 
\cite{colombia}, \cite{turkey}.

As an example of the spinor representation of the Frenet equation we show    
the relation between the time evolution of the spinor (that describes a string)  to the integrable hierarchy of the  nonlinear  Schr\"odinger equation   \cite{faddetak}. 
In particular, it is explained how a   Lax pair that emerges as a consistency condition of the time dependent extension of the spinor Frenet equation, can be chosen so that it coincides with the \underline{su}(2) Lie algebra  valued Lax pair  of the NLSE. 
That way  the conserved  charges of the NLSE hierarchy can be taken as Hamiltonians,  governing  the time evolution and computing  the energy of a  string.
Then  the spinor description into a projection operator formalism is presented.
This kind of operator formalism has been previous utilized extensively, to analyze the integrable  $\mathbb C \mathbb P^{\mathrm N}$ models \cite{berg}-\cite{wojtek}. 
Finally, it is shown how the  conserved charges of the NLSE hierarchy appear from this formalism. 

Next it is shown that  the Lax pair  emerging from the spinor description relates in a natural fashion to the  flat connection of two dimensional SU(2) Yang-Mills theory. 
Then this relation is studied in  terms of a decomposed description of the Yang-Mills theory \cite{fadde1}, \cite{fadde2}.
In addition,  the spinor representation of the Gau\ss-Godazzi equation and its relation  with the decomposed Yang-Mills theory is presented.  
Finally, the occurrence of the NLSE equation and the Liouville equation  when the metric tensor and the second fundamental form are decomposed in a manner that parallels the decomposition of the Yang-Mills connection, is presented.

We conclude by  introducing the  Cartan geometry \cite{cart1}-\cite{wise}. 
This provides a natural framework for  combining the concept of integrability with strings, two dimensional surfaces, and Yang-Mills theories. 
In particular, a decomposed representation of the four dimensional SU(2)
Yang-Mills theory \cite{ludvig}, \cite{chernolud} is investigated and is shown   that it is  a universal theory that governs various  two dimensional integrable models as embedded structures.  
In particular, in terms of several examples,  is is shown that two dimensional integrable models can be  embedded in the  structure of the decomposed  D=4 Yang-Mills.

%
%
%
%
%
%
%
%
%

\section{Classic  Frenet Equation 
}

\subsection{The equation}

We start with a review of the classic
Frenet equation \cite{frenet}, \cite{spivak} that describes the geometry of  a
class $\mathcal C^3$ differentiable string
$\mathbf x(z)$  in $\mathbb R^3$. The parameter $z \in [0,L]$ is  
generic and $L$ is the length of the string in $\mathbb R^3$ defined by
\begin{equation*}
L =  \int\limits_0^L \! dz \,  \sqrt{ {\mathbf x}_z \cdot {\mathbf x}_z } 
\ \equiv \ \int\limits_0^L \! dz \,  \sqrt{ g }.
\end{equation*} 
Note  that this is the time independent 
part of the  Nambu-Goto action.
Let us assume, for simplicity,
that there are no inflection points {\it i.e.} points where the curvature of the string vanishes.
An inflection point is not generic, a simple inflection point can always be
removed by a small generic deformation of the string.

The string can then be globally framed as follows:
The unit length tangent vector 
\begin{equation}
\mathbf t \ = \  \frac{1}{ \sqrt{g} } \frac{ d \hskip 0.2mm \mathbf x (z)} {dz} 
\ \equiv \  \frac{1}{ || {\mathbf x}_z || } \,  
{\mathbf x}_z
\label{curve}
\end{equation}
is orthogonal to the  unit length bi-normal vector
\[
\mathbf b \ = \ \frac{  {\mathbf x}_z\times  {\mathbf x}_{zz} } { || 
 {\mathbf x}_z \times {\mathbf x}_{zz} || },
\]
while the unit length normal vector is given by
\[
\mathbf n = \mathbf b \times \mathbf t.
\] 
Then, the three  
vectors $(\mathbf n, \mathbf b, \mathbf t)$ form the right-handed orthonormal Frenet frame,   at each point of the string. 

The Frenet equation relates the frames at different points along the string  \cite{frenet}, \cite{spivak}. Explicitly, it 
is of the form
\begin{equation}
\frac{d}{dz}\!\left(
\begin{matrix} 
{\bf n} \\
{\bf b} \\
{\bf t} \end{matrix} \right) =  \sqrt{g}  \left( \begin{matrix}
0 & \tau & - \kappa  \\ -\tau & 0 & 0 \\ \kappa & 0 & 0 \end{matrix} \right) 
\left(
\begin{matrix} 
{\bf n} \\
{\bf b} \\
{\bf t} \end{matrix} \right) 
\label{DS1}
\end{equation}
where 
\begin{equation}
\kappa(z) \ = \ \frac{ || {\mathbf x}_z \times {\mathbf x}_{zz} || } { ||  {\mathbf x}_z||^3 }
\label{kg}
\end{equation}
is the curvature of the string on the osculating plane that is spanned by $\mathbf t$ and $\mathbf n$, and
\begin{equation}
\tau(z) \ = \ \frac{ ( {\mathbf x}_z \times  {\mathbf  x}_{zz} ) \cdot { {\mathbf x}_{zzz} }} { || {\mathbf x}_z \times  {\mathbf x}_{zz} ||^2 }
\label{tau}
\end{equation}
is the torsion of the string.

In this case, if the local scale factor $\sqrt{g}$ is determined, 
and the curvature and the torsion   are known, the frames can be constructed by solving (\ref{DS1}).
In addition,   the string can be constructed by solving (\ref{curve}). 
Note that, this solution is unique up to rigid translations and 
rotations of the string. 

Thus, the  Frenet equation construct reparametrization invariant energy functions  in terms of the curvature and the torsion of  the string.

In the following let us  denote by $s \in [0,L]$ the arc-length parameter, 
while $z$ denotes a generic parametrization.
The arc-length parameter $s$  measures
the length along the string in terms of the distance scale of the three dimensional ambient space $\mathbb R^3$.
The change of variables from a generic parameter $z$ to the arc-length parameter $s$ is
\[
s(z) = \int\limits_0^z || {\mathbf x}_z (\tilde z) || d\tilde z.
\]
Accordingly, we consider the effects of infinitesimal local diffeomorphisms 
along the string, obtained by deforming $s$ as follows
\begin{equation}
s \to  z = s + \epsilon(s).
\label{infi}
\end{equation}
Here $\epsilon(s)$ is an arbitrary infinitesimally small function  such that
\[
\epsilon(0) = \epsilon (L) = 0 = \epsilon_s (0)  = \epsilon_s(L).
\]
The Lie algebra of diffeomorphisms (\ref{infi}) of a line segment in $\mathbb R^1$ is the 
classical Virasoro (Witt) algebra.   

Next we define a  function $f(s)$ with support on the string to have a weight $h$ akin to the conformal weight, 
if $f(s)$ transforms according to
\begin{equation}
\delta f (s) = - \left( \epsilon \frac{d}{ds} +  h\epsilon_s \, \right) \, f(s)
\label{fe}
\end{equation}
under the infinitesimal diffeomorphism (\ref{infi}). 
Since the three dimensional geometric shape of the string in $ \mathbb R^3$ does not
depend on the way how it has been parametrized, the embedding 
$\mathbf x(z)$  transforms as a scalar {\it i.e.} it has weight $h=0$
under reparametrizations.
Similarly, the curvature (\ref{kg}) and the torsion (\ref{tau}) are
scalars  under reparametrizations. Infinitesimally, 
\[
\delta \kappa (s) = - \epsilon(s) \frac{d\kappa}{ds} \ \equiv \ -\epsilon {\kappa}_s
\]
\[
\delta \tau (s) = - \epsilon(s) \frac{d\tau}{ds} \ \equiv \ -\epsilon \tau_s.
\]

For the arc-length parameter given by
\[
|| {\mathbf x}_s || \equiv || {\mathbf x}^\prime || =1
\]
 the Frenet equation becomes
\cite{frenet}-\cite{eisenhart}, \cite{hanson}, \cite{kuipers}
\begin{equation}
\frac{d}{ds}\left(
\begin{matrix} 
{\bf n} \\
{\bf b} \\
{\bf t} \end{matrix} \right) =  \left( \begin{matrix}
0 & \tau & -\kappa  \\ -\tau & 0 & 0 \\  \kappa & 0 & 0 \end{matrix} \right) 
\left(
\begin{matrix} 
{\bf n} \\
{\bf b} \\
{\bf t} \end{matrix} \right).
\label{contDS1}
\end{equation}
Note that the framing of the string by the Frenet dreibein is not unique, {\it i.e.}  there are many ways to 
frame a string  \cite{bishop}, \cite{hanson}, \cite{kuipers}.  

Instead of the zweibein ($\mathbf n , \mathbf b$) any two mutually orthogonal vectors ($\mathbf e_1 , \mathbf e_2 $) on the normal planes, {\it i.e.} planes that are perpendicular to the tangent vectors $\mathbf t $, can be chosen.  
Such a generic frame is  related to the Frenet zweibein by a local SO(2) frame rotation around the tangent vector  $\mathbf t(s)$ of the from
\begin{equation}
\left( \begin{matrix} {\bf n} \\ {\bf b} \end{matrix} \right) \ \to \ \left( \begin{matrix} {{\bf e}_1} \\ {\bf e}_2 \end{matrix} \right) \
= \ \left( \begin{matrix} \cos \eta(s) & - \sin \eta(s) \\  \sin \eta(s) & \cos \eta(s) \end{matrix}\right)
\left( \begin{matrix} {\bf n} \\ {\bf b} \end{matrix} \right).
\label{newframe}
\end{equation}
For  the Frenet equation this yields
\begin{equation}
\frac{d}{ds} \left( \begin{matrix} {\bf e}_1 \\ {\bf e }_2 \\ {\bf t} \end{matrix}
\right) =
\left( \begin{matrix} 0 & (\tau - \eta^\prime) & - \kappa \cos \eta \\ 
- (\tau - \eta^\prime)  & 0 & - \kappa \sin \eta \\
\kappa \cos \eta &  \kappa \sin \eta  & 0 \end{matrix} \right)  
\left( \begin{matrix} {\bf e}_1  \\ {\bf e }_2 \\ {\bf t} \end{matrix}
\right).
\label{contso2}
\end{equation}
 
Next let us introduce the three  matrices
\[
T_1 = \left( \begin{matrix} 0 & 0 & 0 \\
0 & 0 & -1 \\ 0 & 1 & 0 \end{matrix} \right), \ \  T_2 = \left( \begin{matrix} 0 & 0 & 1 \\
0 & 0 & 0 \\ -1 & 0 & 0 \end{matrix} \right), \ \  T_3 = \left( \begin{matrix} 0 & -1 & 0 \\
1 & 0 & 0 \\ 0 & 0 & 0 \end{matrix} \right)
\]
that determine the canonical adjoint representation of \underline{so}(3) Lie algebra,
\[
\left[T_a , T_b \right] = \epsilon_{abc} T_c.
\]
Then the action of the SO(2) frame rotation on  $\kappa$ and $\tau$  is described by
\begin{eqnarray}
\hspace{-18mm} \kappa T_2  & \to &
e^{\eta T_3} \left( \kappa\, T_2 \right)  e^{-\eta T_3}  =  \kappa \left( \cos \eta\, T^2 + \sin \eta\, T^1\right),
\label{sok}\\
\hspace{-8mm}\tau \, T_3 & \to & \left(\tau - \eta^\prime \right)  T_3.
\label{sot}
\end{eqnarray}
Observe that (\ref{sok}) and (\ref{sot}) has the same format as the SO(2) $\simeq$ U(1) gauge transformation 
of an Abelian Higgs multiplet. The curvature $\kappa$ is like
a complex valued ``Higgs field" with a real part that
coincides with the $T^2$ components on (\ref{sok}), and an imaginary part that coincides
with the $T^1$ component (or {\it vice versa}). 
Finally, the torsion $\tau$ transforms like the U(1) ``gauge field" of the multiplet.

Let us conclude, by pointing out that the choice  
\begin{equation}
\eta(s) =  \int_0^s \! \tau (\tilde s) d\tilde s 
\label{partran}
\end{equation}
that brings about  the ``$\tau = 0$ " gauge 
yields the parallel transport framing \cite{bishop}. Unlike the Frenet 
framing that can not be defined at an inflection
point of the string, the parallel transport framing can be defined  
continuously and  unambiguously through inflection points and straight segments \cite{hanson}.  

\vskip 0.3cm
\subsection{Time evolution}

Let us  proceed with the generic frame (\ref{sok}) and (\ref{sot}) by setting  
\begin{equation}
\begin{matrix}
\tau_r & = & \tau - \eta^\prime \\
\kappa_n & = & \kappa \sin\eta \\
\kappa_g & = & \kappa \cos\eta.
\end{matrix}
\label{identif}
\end{equation}
Then, in the form of a Darboux {\it tri\`{e}dre}, the Frenet equation (\ref{contso2})  becomes
\begin{eqnarray}
\frac{d}{ds} \left( \begin{matrix} { \bf x } \\ {\bf e}_1 \\ {\bf e }_2 \\ {\bf t} \end{matrix}
\right) & =&  
\left( \begin{matrix} 0 & 0 & 0 & 1 \\
0 & 0  & \tau_r  & - \kappa_g  \\ 
0 & - \tau_r   & 0 & - \kappa_n  \\
0 & \kappa_g   &  \kappa_n    & 0 \end{matrix} \right)  
\left( \begin{matrix}{\bf x} \\  {\bf e}_1  \\ {\bf e }_2 \\ {\bf t} \end{matrix}
\right)\nonumber\\
  &\equiv &  \mathcal R_s \left( \begin{matrix} {\bf x} \\ {\bf e}_1  \\ {\bf e }_2 \\ {\bf t} \end{matrix}
\right).
\label{contso2b}
\end{eqnarray} 
Accordingly, let us assume that  the string $\mathbf x(s)$   lie
on a surface $\mathcal S$ which is embedded in
$\mathbb R^3$. This surface is a
putative world-sheet of the string, swept by its time evolution. 
Then, the vector ${\mathbf e}_2$ can be geometrically interpretred as the unit normal of the surface,
so that $\mathbf t$ and ${\mathbf e}_1$ span its tangent plane at the point $\mathbf x(s)$.
Also,  $\kappa_g$ is the geodesic curvature;  $\kappa_n$ is the normal 
curvature;   and $\tau_r$ is the relative (geodesic) torsion of the string on $\mathcal S$. 
{\bf Remark:} Different choices of $\eta$  in (\ref{contso2}) and (\ref{identif})
correspond to different choices of the surface $\mathcal S$. The surface becomes uniquely determined only once the time evolution of $\mathbf x(s)$  
is specified. 

In analogy with (\ref{newframe}) let us  introduce a U(1) rotation with an angle $\chi$ of the Darboux {\it tri\`{e}dre} with  the rotation around the normal vector $\mathbf e_2$ of the surface $\mathcal S$. Thus,  
it rotates the zweibein of the tangent plane as follows
\[
\left( \begin{matrix}  {\bf e}_1  \\  {\bf t} \end{matrix} \right) \ \buildrel{U(1)}\over{\longrightarrow} \
\left( \begin{matrix}  \cos\chi  & -\sin\chi  \\
\sin\chi  & \cos\chi     \end{matrix} \right)  \left( \begin{matrix}  {\bf e}_1  \\  {\bf t} \end{matrix} \right). 
\]
Then the Frenet equation (\ref{contso2b}) transforms to
\[
\frac{d}{ds} \left( \begin{matrix} { \bf x } \\ {\bf e}_1 \\ {\bf e }_2 \\ {\bf t} \end{matrix}
\right)  \ \buildrel{\chi}\over{\longrightarrow} \   \mathcal R^\chi_s \left( \begin{matrix} {\bf x} \\ {\bf e}_1  \\ {\bf e }_2 \\ {\bf t} \end{matrix}\right)
\]
where the ($ij$) elements of the matrix $\mathcal R_s^\chi$ are given by
\begin{equation*}
\begin{matrix}
\mathcal R_s^\chi (12) & & &=& -\sin\chi \\
\mathcal R_s^\chi (14) & & &=& \cos\chi \\
\mathcal R_s^\chi (23) & = & - \mathcal R_s^\chi(32) &=& \tau_r \cos\chi - \kappa_n \sin\chi \\
\mathcal R_s^\chi (34) & \equiv & - \mathcal R_s^\chi(43) &=& - \tau_r \sin\chi - \kappa_n \cos\chi \\
\mathcal R_s^\chi (24) & = & - \mathcal R_s^\chi(42) &=& - (\kappa_g - \chi^\prime), 
\end{matrix}
\end{equation*}
{\it i.e.} the transformation law of the Abelian Higgs multiplet. 
However, in this context the ``Higgs field"
has as its real and imaginary components the normal curvature $\kappa_n$ and the geodesic torsion $\tau_r$, respectively; while  the geodesic curvature 
$\kappa_g$ is like the ``gauge field". 

Note that after the rotation by $\chi$, the tangent of the string points in the direction is equal to 
\[
{\mathbf x}^\prime(s) \ = \ \cos \chi \, \mathbf t - \sin\chi \, {\mathbf e}_1
\]
while,  in analogy with (\ref{partran}),  in the frame where 
\[
\chi(s) \ = \ \int_0^s \! \kappa_g(\tilde s ) d\tilde s
\]
the geodesic curvature vanishes. Thus, in this frame the direction of the string 
coincides with that of a geodesic on $\mathcal S$.
 
The preceding proposes us to consider the consequences of extending 
$\mathbf x(s)$ into a one parameter 
family of strings $\mathbf x(s,t)$; where 
$t$ is the time so that  $\mathbf x(s,t)$ determines a surface which is
the world sheet $\mathcal S$ of the string.
Then the time evolution transports the Frenet frames (\ref{contDS1}) 
along the direction $\dot {\mathbf x} (s,t)$ on the surface. By completeness of the dreibein, the time evolution 
is governed by an equation of the form
\begin{equation}
\frac{d}{dt} \left( \begin{matrix} {\bf n} \\ {\bf b } \\ {\bf t} \end{matrix}
\right) \ = \ \mathcal R_t \left( \begin{matrix} {\bf n}  \\ {\bf b} \\ {\bf t} \end{matrix}
\right),
\label{contso2c}
\end{equation}
where $\mathcal R_t$ takes values in the \underline{so}(3) Lie algebra,
\begin{equation}
\mathcal R_t \ = \ u T_1 +  v T_2 + w T_3 \ \equiv \ 
\left( \begin{matrix} 0 & -w  & v  \\ 
w  & 0 & -u  \\
-v  &  u   & 0 \end{matrix} \right) .
\label{contso2d}
\end{equation}
Here
\begin{equation*}
\begin{matrix} u & = &  &  \dot{\mathbf t} \cdot {\mathbf b} \\
v & = & & \dot{\mathbf n} \cdot {\mathbf t}  \\
w & = &  & \ \ \dot{\mathbf b} \cdot {\mathbf n}. 
\end{matrix} 
\end{equation*}
Thus  ($ {\bf n}, {\bf b}, {\bf t}$) becomes extended  into an orthonormal 
dreibein, at each point $\mathbf x(s,t)$ of the surface.

For strings $\mathbf x(s,t)$ that are of class $\mathcal C^3$, the linear problem (\ref{contso2b}) and (\ref{contso2c}) is integrable since the ordering of derivatives commute¬
\begin{equation*}
\frac{\partial}{\partial s} \frac{\partial}{\partial t} \ = \ \frac{\partial}{\partial t}\frac{\partial}{\partial s},
\end{equation*}
leading to the zero curvature condition
\begin{equation}
\mathcal F_{ts} \ \equiv \ 
\partial_t \mathcal R_s -  \partial_s \mathcal R_t + [ \, \mathcal R_s \, , \, \mathcal R_t \, ] \ = \ 0,
\label{Fts}
\end{equation}
which corresponds to the Gau\ss-Godazzi equation that describes the
embedding of the surface $\mathbf x(s,t)$ in the ambient   space $\mathbb R^3$  \cite{oma3d}, \cite{foot}. 
Moreover,  we recognize in (\ref{Fts}) the \underline{so}(3) Lax pair structure. 

Indeed,  the Frenet equation together with the time evolution equation (\ref{contso2c}),
admits an interpretation as an auxiliary 
linear problem that has a fundamental role in the theory of integrable models \cite{faddetak}.

Let us conclude by showing how the (focusing) integrable nonlinear Schr\"odinger equation 
\begin{equation}
\frac{1}{i} \partial_t q \ = \  \partial_{ss} q + \frac{1}{2}  | q |^2 q
\label{nlse3d}
\end{equation}
for  $q(s,t)$ complex field,  can be obtained from (\ref{Fts}) in the case of a string that moves under the influence of the local induction approximation (LIA) \cite{darios}-\cite{ricca2}. 
Recall that, the LIA emerges from the Euler equations  as the 
leading order self-induced inextensional motion of a vortex filament in an irrotational fluid \cite{shaffman}.

Let us start  with the expansion \cite{ger}-\cite{ricca1}
\[
\frac{ \partial \, {\mathbf x}(s,t) }{\partial t} \ \equiv  \ \dot{\mathbf x} \ = \ v_t \mathbf t + v_n \mathbf n + v_b \mathbf b
\]
that follows from the completeness 
\begin{equation*}
\partial_s \partial_t \mathbf x = \partial_t \partial_s \mathbf x \ = \ \dot{\mathbf t}.
\end{equation*}
Then the Frenet equation (\ref{contDS1}) take the form 
\begin{equation}
\hskip 0.5cm\begin{matrix} 
u & = & \hskip -4.1cm  \, v_n \tau + v_b' \\
v & = & \hskip -3.1cm - v_t \kappa - v_n' + v_b \tau \\
w & = & \hskip -0.05cm - \frac{1} {\kappa} \left[ \tau ( v_t \kappa + v_n'  - v_b \tau) 
+ (v_n \tau + v_b')' \right]. 
\end{matrix}
\label{riccaeq}
\end{equation}

\noindent
However, the local induction approximation 
\[
\dot{\mathbf x} \ = \ \kappa \mathbf b \ \equiv \ v_b  \mathbf b 
\]
  simplifies   (\ref{riccaeq}) into
\[
\begin{matrix}
u & = &  \kappa ' \\
v & = & \kappa\tau \\
w & = &  \tau^2 - \frac{1}{\kappa} \kappa'' .\\
\end{matrix}
\]
By  demanding that the Lax pair ($\mathcal R_s, \mathcal R_t$) obeys (\ref{Fts}) {\it i.e.} derivatives commute, the following  conditions are derived
\begin{equation}
\begin{matrix}
\dot \kappa & = & - (\kappa \tau)' - \kappa'\tau \\
\dot \tau & = & \left( \frac{\kappa'' - \kappa \tau^2}{\kappa}\right)' + \kappa' \kappa.
\end{matrix}
\label{darios}
\end{equation}
Finally, by introducing the Hasimoto variable \cite{hasimoto}
\begin{equation*}
q(s,t) \ = \ \kappa(s,t) e^{ i \int^s\! \tau(u,t) du}
\end{equation*}
the zero curvature condition (\ref{darios}) becomes equivalent to (\ref{nlse3d}).

%
%
%
%
%
%
%

%
%
%
%
%
%
%
%
%
%
%
%
%
%

\section{Spinor Frenet Equation 
}

\noindent
In this section,  the classic Frenet equation of a string is described in terms of spinors. Let us  start by introducing  a two complex component spinor  along the string {\it i.e.}
\begin{equation}
\psi: \, [0,L] \to \mathbb C^2  \  ; \ \ \ \  \psi =  \left( \begin{matrix} z_1  \\ z_2 \end{matrix}\right). 
\label{psi1}
\end{equation}
In addition, the conjugate spinor $\bar\psi$ is  obtained by acting 
 the ``charge conjugation'' operation $\mathcal C$ on $\psi$:
\begin{equation}
\mathcal C \psi \ = \  -i \sigma_2 \psi^\star = \bar \psi,
\label{C}
\end{equation}
where   $\hat \sigma = (\sigma_1, \sigma_2, \sigma_3)$ are the standard Pauli matrices.
In terms of the spinor components, the charge conjugation amounts to
\[
z_\alpha \ \buildrel{\mathcal C}\over{\longrightarrow}  \, - \epsilon_{\alpha \beta} 
\,
z_\beta^\star
\]
where $\epsilon_{12} = 1$ and $ \epsilon_{\alpha \beta} = - \epsilon_{\alpha \beta}$.
Also,  note that
\[
\mathcal C^2 = - \mathbb I.
\]
Next, we introduce the normalization  condition
\begin{equation}
\psi^\dagger \psi  \ \equiv \ <\psi,\psi> = 1 = <\bar\psi , \bar\psi> = \bar\psi^\dagger \bar\psi
\label{ortho}
\end{equation}
so that $\psi$ maps the string onto a complex two-sphere; in accordance with the orthonormality condition
\[
<\psi \, , \bar\psi> = 0.
\]
Thus, the following completeness exists
\[
|\psi><\psi | + |\bar\psi > <\bar\psi | = \psi_\alpha \psi_\beta^\dagger + \bar\psi_\alpha \bar\psi_\beta
^\dagger \ = \ \mathbb I_{\alpha\beta}
\]
 leading to the differential equation 
\begin{equation}
\partial_s \psi \ = \  <\psi ,  \partial_s \psi> \! \psi \ + <\bar \psi , \partial_s \psi> \! \bar \psi. 
\label{sfrenet}
\end{equation}
By defining  the complex valued curvature as
\begin{equation}
\kappa_c  \, = \, \kappa_g  + i \kappa_n \ = \ 2\! <\bar\psi , \partial_s \psi > 
\label{kappar}
\end{equation}
and  the real valued torsion as
\begin{equation}
\tau_r \, = \,  2  i <\psi, \partial_s \psi> 
\label{taur}
\end{equation}
equation (\ref{sfrenet}) transforms to 
\begin{equation}
\partial_s \psi \ = \  - \frac{i}{2} \tau_r  \psi +  \frac{\kappa_c}{2} \bar \psi,
\label{frenet1}
\end{equation}
the so-called  \emph{spinor Frenet equation}.
So, for given  ($\kappa_c\, ,\tau_r$), the spinor $\psi$ can be constructed along the string. 
Additionally, from the spinor  $\psi$ the unit length vector 
\begin{equation}
{\mathbf t} = <\psi, \hat \sigma \psi> \ = \ - <\bar\psi , \hat \sigma \bar\psi>
\label{tpsii}
\end{equation}
can be obtained in order to  evaluate the string $\mathbf x(s)$, by identifying $s$ as
the proper-length parameter and by integrating
\begin{equation}
 \frac{ d \mathbf x}{ds} = {\mathbf t} (s) \ \equiv \ <\psi, \hat \sigma \psi>.
\label{txpsi}
\end{equation}
Thus,   the string is determined uniquely up to a global rotation and a global translation.

The normalization condition (\ref{ortho}) can be relaxed. By assuming  
\[
<\psi,\psi> = <\bar\psi , \bar\psi> = \sqrt{\mathfrak g(s)}
\]
the unit length vector takes the form
\[
 \mathbf t \ = \ \frac{1}{\sqrt{\mathfrak g(s)}}
 <\psi, \hat \sigma \psi> .
\]
By comparing   with (\ref{curve}) the following  correspondence occurs
\[
\sqrt{\mathfrak g} \simeq || \mathbf x_z || = \sqrt{g}.
\]
Thus, re-defining the normalization of the spinors by
\[
|\psi> \to (\mathfrak{g})^{-1/4} |\psi>
\]
which effectively sends $\sqrt{\mathfrak g} \to 1$,
corresponds to the choice of arc-length parameterization along the string. 

%
%
%
%
%
%
%
%
%
%
%

\section{Frenet frames from spinors}

In what follows,  the relation between the classic  version and the spinor version of the Frenet equation is studied in detail. 
Consider the  local U(1) rotation
\begin{equation}
\psi \to e^{i\varphi} \psi, \ \ \ \ \\ \ \ \ \bar\psi \to e^{-i\varphi} \bar\psi.
\label{u11}
\end{equation}
Then  the complex curvature defined in (\ref{kappar})  becomes
\begin{equation}
\kappa_c \to e^{2i\varphi} \kappa_c
\label{u1kappa}
\end{equation}
while  the torsion (\ref{taur}) transforms to
\begin{equation}
\tau_r \ \buildrel{\varphi}\over{\longrightarrow} \  \tau_r -  \, 2 \, \partial_s \varphi.
\label{u1tau}
\end{equation}
In this case, both the vector $\mathbf t$ defined by (\ref{tpsii}) and the string $\mathbf x(s)$ obtained from (\ref{txpsi})
  remain invariant under the U(1) rotation (\ref{u11})-(\ref{u1tau}).

In addition of $\mathbf t$, let us  introduce the  two complex vectors
\begin{eqnarray}
\hspace{-5mm}{\mathbf  e}_+ & = & \frac{1}{2} (
{\mathbf e}_1 + i \, {\mathbf e}_2 ) \ = \, 
\frac{1}{2} \!  <\bar \psi , \hat \sigma \psi >   \ \equiv \ \frac{1}{2}  {\bar \psi}^\dagger \hat\sigma \psi
\label{e+}\\
\hspace{-9mm}{\mathbf  e}_-  &=&  \frac{1}{2} (
{\mathbf e}_1 - i \, {\mathbf e}_2) \  = \, 
\frac{1}{2} \! < \psi , \hat \sigma \bar \psi >   \ \equiv \ \frac{1}{2}  {\psi}^\dagger \hat\sigma \bar\psi
\label{e-}
\end{eqnarray}
where  ${\mathbf e}_1$ and ${\mathbf e}_2$ are the real and imaginary parts of
$\mathbf e_\pm$, respectively.  A direct computation shows that
\[ 
{\mathbf e}_i \cdot 
{\mathbf e}_j = \delta_{ij}, \ \ \ \ \ \ \ \ \ \ \ {\mathbf e}_i \cdot {\mathbf t} = 0
\]
and 
\[
\left({\mathbf e}_1 \times {\mathbf e}_2\right) \cdot \mathbf t = 1.
\]
Thus (${\mathbf e}_1, {\mathbf e}_2 , {\mathbf t}$) is a right-handed 
orthonormal system. In particular,  (${\mathbf e}_1 , {\mathbf e}_2 $) span
the normal planes of the string that is obtained by integration of (\ref{txpsi}). 

The local U(1) rotation (\ref{u11}) brings about the frame rotation
\[
{\mathbf e}_\pm  \ \to \ e^{\pm 2i\varphi} {\mathbf e}_\pm
\]
and a direct computation using the definitions 
(\ref{kappar}), (\ref{taur}), (\ref{e+}) and  (\ref{e-}) leads to the following system
\begin{equation}
 \begin{matrix} \partial_s  {\bf e}_+  & = & \\  \\
{\partial_s \mathbf t }   & = & \end{matrix}
\begin{matrix}  - i \tau_r \mathbf e_+ - \frac{1}{2} \kappa_c \mathbf t \\ \\
4 \left( \kappa_c  \mathbf e_+ +  \kappa_c^\star \mathbf e_- \right).
\end{matrix}  
\label{contso3}
\end{equation} 
This coincides with  the general frame (Darboux) equation  (\ref{contso2b}) when 
$\kappa_c$ is given by (\ref{kappar}). 
Note that different choices of $\varphi$ correspond
to different choices of embedding surfaces.

To complete the interpretation of the Frenet frames (\ref{contDS1}) in terms of spinors,
introduce the following (local) coordinate representation  of the components of spinor (\ref{psi1}):
\begin{equation}
z_1 =  \cos\frac{\vartheta}{2} \, e^{i\phi_{1}},  \ \ \ \ \  \  \ \ \ \ \
z_2 = \sin\frac{\vartheta}{2} \,  e^{i\phi_{2}}.
\label{para}
\end{equation}
Then the tangent vector of the string (\ref{tpsii}) using (\ref{para}) gives  
\begin{equation}
\mathbf t \ = \ \left( \begin{matrix} \cos \phi_- \sin \vartheta \\ \sin\phi_- \sin \vartheta \\
\cos \vartheta \end{matrix} \right)
\label{texp}
\end{equation}
where 
\[
 \phi_\pm = \phi_2 \pm \phi_1.
\]
In addition, by introducing the unit vectors
\begin{equation}
\mathbf u \ = \ \left( \begin{matrix} \cos \phi_- \cos \vartheta \\ \sin\phi_- \cos \vartheta \\
- \sin\vartheta \end{matrix} \right),  \ \ \ \ \ \ \ 
\mathbf v \ = \ \left( \begin{matrix} -\sin \phi_-  \\ \cos \phi_- \\ 0 
\end{matrix} \right)
\label{uv}
\end{equation}
and  substituting (\ref{para}) in (\ref{e+}) and  (\ref{e-})  one obtains
\begin{equation}
\mathbf e_\pm \ = \ \frac{1}{2} e^{\pm i\phi_+} ( \mathbf u \pm i \mathbf v).
\label{e+-1}
\end{equation}
Furthermore,  (\ref{texp}) becomes
\[
\partial_s \mathbf  t  \ = \  \partial_s \vartheta \, \mathbf u \,
+ \, \sin \vartheta  \, \partial_s \phi_-  \, \mathbf v. 
\]
while,  similarly to (\ref{contDS1}), one gets
\begin{equation}
\left(\partial_s \vartheta \right) \, \mathbf u \,
+ \, \sin \vartheta  \, \partial_s \phi_-  \, \mathbf v 
 \ = \ \kappa \mathbf n.
\label{un}
\end{equation}
Here $\kappa$ is the (geometric Frenet) curvature and 
$\mathbf n$ is the normal vector of the string while,
\[
\kappa^2  \ = \ \kappa_c \kappa_c^\star \ = \  \kappa_g^2 + \kappa_n^2  \ = \  
\left(\partial_s \vartheta \right)^2 + \sin^2\! \vartheta  \left(\partial_s \phi_-\right)^2.
\]
On the other hand, using the complex curvature (\ref{kappar}) in   the parametrization (\ref{para}) leads to
\begin{equation}
\kappa_c =   2\left(z_1 \partial_s z_2 - z_2 \partial_s z_1 \right)
= \ e^{i \phi_+} \left( \partial_s \vartheta + i \, \sin \vartheta \, \partial_s \phi_-  \right). 
\label{kappaphi}
\end{equation}

Due to    (\ref{kappar}) and  (\ref{contso3}) the Frenet framing of the string is  specified by demanding that $\kappa_c$ is real. Thus,
\[
<\bar\psi, \partial_s \psi> - <\psi, \partial_s \bar\psi> \ = \ 0.
\]
In terms of (\ref{para}) this reads
\[
\sin \phi_+ \frac{d \vartheta }{ds} \, +  \, \cos \phi_+ \sin\vartheta \, \frac{d\phi_-}{ds\,} 
\ = \ 0.
\]

The U(1) rotation (\ref{u11}) with the choice
\begin{equation}
2\varphi = -\phi_+
\label{2phi}
\end{equation}
implies that  (\ref{kappaphi}) transforms to
\[
\kappa_c \, \to  \, 
\partial_s \vartheta + i \, \sin \vartheta \, \partial_s \phi_-  
\]
and due to the choice of the phase $\phi$ one obtains that 
\[
 \kappa_g \simeq \partial_s \vartheta, \ \ \ \ \ \ \ \ \ \ \ \ \
\kappa_n \simeq \sin\!\vartheta \,\partial_s \phi_-.
\]
This can be interpreted as follows: Take $\vartheta(s)$ and $\phi_-(s)$ to be the 
coordinates on a two-sphere $\mathbb S^2_R$ with radius $R$. Assume 
this sphere osculates the string at the point
$\mathbf x(s)$,  in such a manner that the tangent of the string becomes
parallel with the tangent of a  great circle of the sphere.  
At the point of contact the value of
$R$ then coincides with the inverse 
of the curvature $\kappa(s)$ of the string. 
However, the great circle is a geodesic of $\mathbb S^2_R$ and $s$ 
is the proper length parameter so
\[
\frac{1}{R^2} = (\partial_s \vartheta)^2 + \sin^2\! \vartheta \, 
(\partial_s \phi_-)^2  \ = \ \kappa^2.
\]
By orienting the osculating two-sphere so that the 
osculating great circle coincides with the equator $\vartheta=\pi/2$  of the sphere,  the vectors (\ref{uv}) coincide  with the Frenet frame normal and bi-normal vectors, respectively. In particular,
 along the equator  $\vartheta = \pi/2$ (is constant) so  (\ref{un}) implies
\begin{equation}
 \partial_s \phi_- \, \mathbf u \ = \ \kappa \mathbf n \ = \ \frac{1}{R} \mathbf n \equiv \ \kappa_n \mathbf n
\label{equat}
\end{equation}
since $\kappa_g = 0$ for a geodesic. 

In addition,  the torsion (\ref{taur}) 
\[
\frac{1}{2i}  \, \tau_r \, = \, <\psi, \partial_s \psi> 
= z_1^\star \partial_s z_1 + z_2^\star \partial_s z_2 
\]
 in terms of the parametrization (\ref{para}) becomes
\begin{equation}
\tau_r \, = \,   \cos\vartheta \, \partial_s\phi_-  -  \partial_s \phi_+
 \label{tauphi}
 \end{equation}
while  the U(1) rotation (\ref{2phi}) gives
\[
\tau_r  \, = \,   \cos\vartheta \, \partial_s\phi_-.  
\]
Thus, for  a path along the equator (\ref{equat}) there is no torsion. 
With this choice of  phase $\varphi$, the  torsion $\tau_r$ 
coincides with the U(1) invariant ``super-current"
\begin{equation}
 j = \tau_r - \frac{i}{2} \left\{ \frac{z_1^\star \partial_s z_1 - z_1 \partial_s z_1^\star
 }{z_1z_1^\star} 
 + \frac{ z_2^\star \partial_s z_2 - z_2 \partial_s z_2^\star }{z_2z_2^\star} 
 \right\}.
\label{superj}
\end{equation}
As a consequence it appears natural to identify $j$ with the Frenet frame
torsion $\tau$ of the string.

%
%
%
%
%
%
%
%
%
%
%
%
%
%

\section{Maurer-Cartan relations}

Let us start by noting that  (\ref{kappaphi}) can be interpreted  as
the pull-back  of the complex valued one-form 
\begin{equation}
\kappa_c \ = \   i \mathfrak b_- \ = \  
e^{i \phi_+} \, (\, d \vartheta + i \, \sin \vartheta \, d \phi_- ) 
\label{calb}
\end{equation}
to the surface $\mathcal S$ that is determined by  $\mathbf x(s,t)$. 
Similarly,  (\ref{tauphi}) can be interpreted
as the pull-back of the real valued one-form
\begin{equation}
\tau_r \ = \ - \mathfrak a \ = \ \cos\vartheta \, d\phi_- - d\phi_+
\label{cala}
\end{equation}
to this surface. Equation (\ref{cala}) can be identified as 
the Dirac monopole connection one-form, 
by interpreting ($\vartheta, \phi_-$) as 
the local coordinates of the base $\mathbb S^2$ 
and $\phi_+$ as the coordinate of the U(1) fibre. 

Let $g $ be the SU(2) matrix defined as
\begin{equation}
g = \left( \begin{matrix} z_1 & -\bar z_2 \\ z_2 & \bar z_1 \end{matrix} \right)
\label{g}
\end{equation}
such that
\begin{eqnarray}
\hspace{-10mm}g \sigma_3 g^{-1} &=&{ \mathbf t} \cdot {\vector\sigma} \ \equiv \ \hat {\vector t}
\label{gmat1}
\\
\hspace{-15mm}g \, \sigma_\pm g^{-1} &\equiv & \frac{1}{2} g \left(\sigma_1 \pm i \sigma_2 \right)  g^{-1}
\ = \mathbf e_\pm \cdot {\vector \sigma} \ \equiv \ \hat{\vector e}_\pm.
\label{gmat2}
\end{eqnarray}
Then, the matrices ($\hat{\vector e}_\pm, \hat{\vector t}$) obey the \underline{su}(2) Lie algebra
\begin{equation}
[ \, \hat{\vector t} \, , \, \hat{\vector e}_\pm \, ] \ = \  \pm 2 \hat{\vector e}_\pm, 
\ \ \ \ \ \ \ \ \ \ \ \ 
[ \, \hat{\vector e}_+ \, , \, \hat{\vector e}_- \, ] \ = \   \hat{\vector t}.
\label{su2}
\end{equation}
Note that the matrix  $\hat{\vector t}$  is defined up to a $h\in$ U(1) $\!\!\subset\!$ SU(2) multiplication  of $g$ from the right.  {\it I.e.} for
\begin{equation}
g \ \buildrel h \over \longrightarrow \ gh,
\label{gh}
 \hspace{5mm}
h \, = \, e^{i\varphi \,\sigma_3}
\end{equation}
one obtains that 
\begin{equation}
\hat{\vector t} \ \buildrel h \over \longrightarrow \ g h \sigma_3 h^{-1} g^{-1} 
\equiv \hat{\vector t}.
\label{3nro}
\end{equation}
In addition,
\begin{equation}
\hat{\vector e}_{\pm} \ \buildrel h \over \longrightarrow \
gh\sigma^\pm h^{-1} g^{-1} \ = \ e^{\pm 2i\varphi} \hat{\vector e}_\pm.
\label{3ero}
\end{equation}
Note that (\ref{3ero}) coincides in form with (\ref{e+-1}), by identifying
\[
\phi_+ \simeq 2 \varphi.
\]

Next, let us arrange the components of the 
vectors $(\mathbf  e_\pm, \mathbf t)$ 
into elements of a SO(3) matrix {\it i.e.}
\begin{equation*}
{ {\mathcal O}_i}^a \ \buildrel {\tt def} \over = \
(e^a_1, e^a_2, t^a)
\end{equation*}
which relates the local basis  of the Lie algebra at unity to the Lie algebra basis at the point $g$ on SU(2) since
\begin{equation*}
(\sigma_\pm , \sigma_3) \ \buildrel g \over \longrightarrow
\ {{\mathcal O}_i}^a \sigma_a\ = \ ( \hat{\vector e}_\pm , \hat{\vector t}\,).
\end{equation*}
But  ${{\mathcal O}_i}^a$ define the components of the dreibein, thus  the 
corresponding Levi-Civita connection one-form ${\omega_i}^j$ is subject to
the Cartan structure  equation
\begin{equation*}
d {{\mathcal  O}_i } + {{\omega}_i}^j {\mathcal O}_j = 0.
\end{equation*}
In terms of (${\mathbf e}_{\pm} , \, \mathbf  t $) this becomes 
\[
d {\mathbf e}_{\pm} = \, <\! \mathbf  t , d \mathbf e_\pm \!> \mathbf t + 
2 \! < \! \mathbf e_{\pm} , d \mathbf  e_\pm \! > \mathbf  e_\pm
\]
\[
d \mathbf t = - 2 \! <\! \mathbf t  , d \mathbf  e_+ \!> \mathbf  e_- -  
2 \! < \! \mathbf t , d \mathbf  e_- \! > \! \mathbf e_+.
\]
By  identifying  
\begin{eqnarray}
\hspace{-5mm} {\mathfrak a}\!\! & = &\!\! - 2i \! <\!\mathbf  e_+,
d \mathbf  e_+\! >  = -i  \, \mbox{tr} \left(\sigma_3 g^{-1} 
d g \right) =\!  -i  \, \mbox{tr} \left( \sigma_3 R \right)\nonumber\\
\label{3aco}
\\
\hspace{-3mm} \mathfrak b_{\pm} \!\!&=& \!\!\pm 2i \!\!
 \, <\! \mathbf e_{\pm} , 
d \mathbf t \! >
=\!  - 2i \mbox{tr} \left ( \sigma_\mp g^{-1} d g  \right)= 
\! - 2i  \mbox{tr} \left( \sigma_\mp R \right)\nonumber\\
\label{3bco}
\end{eqnarray}
where  
\begin{equation}
R = g^{-1}  d g
\label{R}
\end{equation}
is the connection one-form, one gets  the (right) Maurer-Cartan form
\begin{equation}
dR + [R,R] = 0.
\label{flatR}
\end{equation}
 In terms of the local coordinates (\ref{para})
we confirm that (\ref{3bco}) and (\ref{3aco})  coincide with the
one-forms  (\ref{calb}) and (\ref{cala}), respectively.
In particular, the U(1) rotation (\ref{3ero}) and  (\ref{3nro})
reproduced (\ref{u1kappa}) and  (\ref{u1tau}) since
\begin{equation}
\begin{array} {ccc}
\mathfrak b_{\pm} \ & \buildrel h \over \longrightarrow
\ & e^{\mp 2i\varphi} \mathfrak b_{\pm} \\
\mathfrak a \ & \buildrel h \over \longrightarrow \
& \mathfrak a + 2 d \varphi.
\end{array}
\label{3abr}
\end{equation}
Thus  $\mathfrak a$ and $\mathfrak b_\pm$  can be combined into  the single SU(2) Maurer-Cartan form $R$ by setting
\begin{equation}
-2i R \ = \  - 2 i \, g^{-1} d  g \ = \   \mathfrak a \, \sigma_3 \, + 
\mathfrak b_{+} \sigma_+   + \mathfrak b_{-}  \sigma_- .
\label{dgg}
\end{equation}

Alternatively, in terms of the (left) Maurer-Cartan form
\begin{equation}
L = d g \, g^{-1} \ = \ g R g^{-1}
\label{L}
\end{equation}
one gets
\begin{equation*}
dL + [L,L] = 0
\end{equation*}
where
\begin{equation}
-2i L = \mathfrak a \, \hat{\mathbf t} + \mathfrak b_{+} \hat{\mathbf e}_+  +  \mathfrak b_{-}  \hat{\mathbf e}_-
\ = \ \mathfrak a \, \hat{\mathbf t} \, + \, \frac{1}{2i} [ \, d \hat{\mathbf t} \, , \, \hat{\mathbf t}\, ].
\label{Lte}
\end{equation}

\vskip 0.5cm

%
%
%
%
%
%
%

%
%
%
%
%
%
%

%
%
%
%
%
%

\vskip 0.5cm

\section{Lax Pair}

\subsection{Non-linear Schr\"odinger equation (NLSE)}

In this section,  the construction of the Lax pair for the nonlinear Schr\"odinger equation \cite{faddetak}
\begin{equation}
\frac{1}{i} \partial_t q \ = \ \partial_{ss} q - 2\lambda |q|^2 q
\label{nlsef}
\end{equation}
is reviewed and then its connection with  the Frenet equation is presented. 

The NLSE Hamiltonian is
\begin{equation}
H_3 \ = \ \int \!ds \,  \left(\left| \frac{ \partial q}{\partial s } \right|^2 + \lambda |q|^4 \right)
\label{nlse}
\end{equation}
where $q$ is a complex variable. The Poisson brackets that determine the time evolution 
are
\begin{eqnarray*}
&&\{ q(s) , q(s') \} \ = \ \{ \bar q(s) , \bar q(s') \} \ = \ 0\\
&&\{ q(s) , \bar q(s') \} \ = \ i \delta(s-s').
\end{eqnarray*}
The NLSE is an integrable equation and its  Lax pair ($U,V$) is defined by
\begin{equation}
U \ = \ U_0 + \xi U_1
\label{U}
\end{equation}
where $\xi$ is a complex spectral parameter and 
\begin{eqnarray}
&& U_0\ = \  \sqrt{\lambda} \left( \begin{matrix} 0 & \bar q \\ q & 0 \end{matrix} \right) \ = \ \sqrt{\lambda}
\left( \bar q \sigma_+ + q \sigma_-\right)
\label{fadU0}
\\
&& U_1\  = \ \frac{1}{2i}  \left( \begin{matrix} 1 & 0 \\ 0 & -1 \end{matrix} \right) \ = \ \frac{1}{2i}\sigma^3.
\label{fadU1}
\end{eqnarray}
In addition,
\[
V \ = \ V_0 + \xi V_1 + \xi^2 V_2
\]
where
\begin{eqnarray}
&&V_0\ = \  i \sqrt{\lambda} \left( \begin{matrix} \sqrt{\lambda} |q|^2  & - \partial_s \bar q 
\\ \partial_s q & - \sqrt{\lambda} |q|^2  \end{matrix} \right)
\nonumber\\
&&\hspace{5mm} \ =\ i  \lambda |q|^2 \sigma_3 - i\sqrt{\lambda} \left( \partial_s \bar q \, \sigma_+ - \partial_s q \, \sigma_- \right)
\label{fadV0}
\end{eqnarray}
and 
\begin{equation}
V_1  \ = \ - U_0, \ \ \ \ \ \ \ \ \ \ \  V_2  \ = \  -U_1.
\label{fadV12}
\end{equation}

The integrability of NLSE is  an outcome of the  over-determinacy of the auxiliary linear equations
\begin{eqnarray}
\frac{\partial}{\partial s} \left( \begin{matrix} f_1 \\ f_2 \end{matrix} \right)  &= & U(s,t,\lambda) 
\left( \begin{matrix} f_1 \\ f_2 \end{matrix} \right) 
\label{aux1}
\\
\frac{\partial}{\partial t} \left( \begin{matrix} f_1 \\ f_2 \end{matrix} \right)  &= & V(s,t,\lambda) 
\left( \begin{matrix} f_1 \\ f_2 \end{matrix} \right)
\label{aux2}
\end{eqnarray}
due to the compatibility condition
\[
\frac{\partial}{\partial s} \frac{\partial}{\partial t} \left( \begin{matrix} f_1 \\ f_2 \end{matrix} \right) 
\ = \ \frac{\partial}{\partial t} \frac{\partial}{\partial s}\left( \begin{matrix} f_1 \\ f_2 \end{matrix} \right)
\]
provided that the Lax pair ($U,V$) comprises the components of a
flat SU(2) connection
\begin{equation}
F_{ts} =  \partial_t U - \partial_s V + [U,V] \ = \ 0.
\label{Fnlse}
\end{equation}
See also (\ref{Fts}).

To sum up, when   the explicit representations
(\ref{fadU0})-(\ref{fadV12}) are substituted in (\ref{Fnlse})  the NLSE (\ref{nlsef}) occurs; while when 
$q$ obeys the NLSE, the Lax pair ($U,V$) is a flat SU(2) connection.

\subsection{Majorana realization of Frenet equation}

Next,  the relation between  the Lax pair ($U,V$)  of the NLSE and the spinor Frenet  equation (\ref{frenet1}) is studied. 
To do so  the spinor Frenet  equation is combined with its  conjugate equation
\begin{equation}
\partial_s \bar \psi = \frac{i}{2} \tau_r \bar\psi - \frac{\kappa_c^\star}{2} \psi
\label{frenet2}
\end{equation}
into a single equation.
This can be done by merging the two spinors into a four-component (conjugate) spinor of the form
\begin{equation}
\Psi = \left( \begin{matrix} - \bar \psi \\ \psi \end{matrix} \right). 
\label{majspi}
\end{equation}
This four-spinor is subject to the Majorana condition: {\it i.e.} under the charge conjugation (\ref{C}) transforms as
\[
\mathcal C \Psi \ = \  \left( \begin{matrix} \psi \\ \bar \psi \end{matrix} \right)  \ = \ i \sigma_2 \Psi \ \equiv \ 
\left( \begin{matrix} 0 & 1  \\ - 1 & 0 \end{matrix} \right) \left( \begin{matrix} - \bar \psi \\  \psi \end{matrix} \right) .
\]
Here the Pauli matrices  $\sigma_a$  act in the two dimensional
space of  the spinor components of $\Psi$. 
In terms of the Majorana spinor $\Psi$, the spinor Frenet equations (\ref{frenet1}) and (\ref{frenet2}) combine to 
\[
\partial_s \Psi \ = \  \frac{i}{2}\left(   \tau_r \sigma_3 -i  \kappa^\star_c \sigma^+ + i  \kappa_c 
\sigma^-\right) \Psi.
\]
By using the  relations (\ref{calb}), (\ref{cala}) and the 
identification Maurer-Cartan one-form (\ref{dgg}) this becomes
\begin{equation}
( \partial_s  +   g^{-1} \partial_s g)  \Psi \ \equiv \ 
( \partial_s   +   R_s )  \Psi \ = \ 0 
\label{Psis}
\end{equation}
where $R_s$ denotes the $s$-component of the right Maurer-Cartan form (\ref{R}).
Observe that in terms of  the SU(2) transformed spinor
\begin{equation}
\Psi_g \ = \ g^{-1} \Psi
\label{gPsi}
\end{equation}
the  equation (\ref{Psis}) changes into
\[
\partial_s \Psi_g \ = \ 0.
\]
Thus the spinor $\Psi_g(s)$ is constant along the string $\mathbf x(s)$.

On the tangent vector (\ref{tpsii}) the SU(2) rotation (\ref{gPsi})  acts as follows  
\[
\mathbf t  \ = \ \left( \begin{matrix} \cos \phi_- \sin \vartheta \\ \sin\phi_- \sin \vartheta \\
\cos \vartheta \end{matrix} \right) \  \buildrel {g^{-1}} \over {\longrightarrow} \left( \begin{matrix} 0 \\ 0 \\
1 \end{matrix} \right).
\]
This means that   (\ref{gPsi}) determines a spatial SO(3) frame rotation, which 
at each point $\mathbf x(s)$  of  the string orients the Cartesian coordinates so that 
the tangent vector points along 
the positive $z$-axis. 
Similarly, the vectors (\ref{uv}) that span the 
normal plane of the string transform like
\begin{equation} 
\mathbf u + i \mathbf v  \buildrel {g^{-1}} \over {\longrightarrow} e^{i\varphi}  \left( \begin{matrix} 1 \\ i \\
0 \end{matrix} \right).
\label{localframe}
\end{equation}
Here the choice  of $\varphi(s)$  specifies a frame on the normal planes. 
If the local frame is chosen so that the direction of the vector $\mathbf u$ in (\ref{localframe}) at each point coincides with the direction of the Frenet frame normal vector along the string, the continuum version of the co-moving framing  is obtained \cite{hu}.

\subsection{Lax pair for the Frenet equation}

Following  (\ref{contso2c}), a one parameter family of string 
$\mathbf x(s,t)$ is considered  with $t$ being time so that $\mathbf x(s,t)$ describes the 
time evolution of the string. By completeness, in terms of the Majorana spinor (\ref{majspi}),  its time evolution is described by the equation 
\begin{equation}
\left( \partial_t +   R_t  \right)  \Psi \ = \ 0. 
\label{Psit}
\end{equation}
Here $R_t (s,t)$ is \underline{su}(2) Lie algebra valued {\it i.e.} it is
a linear combination of Pauli matrices $\sigma_a$ of the form
\begin{equation*}
R_t (s,t) \ = \ \alpha(s,t)  \sigma_1 + \beta(s,t) \sigma_2 + \gamma(s,t) \sigma_3
\end{equation*}
where ($\alpha,\beta,\gamma$) are coefficients. 

In analogy with (\ref{aux1}) and (\ref{aux2}) the linear system  (\ref{Psis}) and (\ref{Psit})  is  over-determined. Its integrability yields to the zero-curvature condition 
\begin{equation}
F_{ts} \ \equiv \ \partial_t R_s - \partial_s R_t + [ \, R_s \, , \, R_t \, ] \ = \ 0 
\label{Fts2}
\end{equation}
implying that  ($R_s, \, R_t$) is a flat two dimensional SU(2) connection one-form. 
Note that  ($R_s, R_t$) can be considered as  the  Lax pair of the spinor Frenet equation; {\it i.e.} the
Gau\ss-Codazzi equation that governs the embedding of the surface $\mathbf x(s,t)$ in the  ambient $\mathbb R^3$.

\subsection{Relation between NLSE and Frenet  Lax pairs}

In what follows,  solution of equation (\ref{Fts2}) is constructed which relates the string and the NLSE. 
Due to  (\ref{calb}) and (\ref{cala})   the relative torsion and the complex curvature can be combined into   the (putative) right Maurer-Cartan form as
\begin{equation} 
-2iR_s \ = \ \tau_r \sigma_3 + i\kappa_c \sigma_+ -i \kappa^\star_c \sigma_-.
\label{Rlax1}
\end{equation}
Next introduce the U(1) transformation {\it a.k.a.} frame rotation of the string (\ref{gh}) which acts on  $R_s$ by sending
\begin{equation}
\begin{array} {ccc}
\kappa_c \ & \buildrel h \over \longrightarrow
\ & e^{- 2i\varphi} \kappa_c \\
\tau_r \ & \buildrel h \over \longrightarrow \
& \tau_r + 2 \partial_s \varphi.
\end{array}
\label{3abr2}
\end{equation}
Here  $\varphi$ is chosen
\begin{equation}
\varphi \ = \ - \frac{1}{2} \int_0^s \! \! ds' \left\{ \tau_r (s') + \xi \right\}
\label{partran2}
\end{equation}
such that  the first term implements the gauge transformation  {\it i.e.} frame rotation
from the generic Darboux frame to the parallel transport frame  (\ref{partran}); while  $\xi$ is the putative complex  spectral parameter.
  
Then the $U(1)$  transformation (\ref{3abr2}) sends 
\[
R_s \  \buildrel h \over \longrightarrow \ R^h_s 
 \]
 where
 \begin{equation*}
R^h_s = \frac{\xi}{2i}  \sigma^3 - \frac{\kappa_c}{2} e^{i \int\limits^s \! \{ \tau_r (s') + \xi \} }
\sigma_+ \, + \, \frac{\kappa^\star_c}{2} e^{-i \int\limits^s \! \{ \tau_r (s') + \xi \} }
\sigma_-.
\end{equation*}
By introducing  the Hasimoto variable
\begin{equation}
 \sqrt{\lambda} \, \bar q  \ = \
 - \frac{\kappa_c}{2} e^{i \int\limits^s \! \{ \tau_r (s') + \xi \} }
\label{kappaide}
\end{equation}
one obtains
\begin{equation}
R^h_s \ = \ 
\frac{\xi}{2i} \sigma^3 + \sqrt{\lambda} \, \bar q \, 
\sigma_+ \, + \, \sqrt{\lambda} \, q\,
\sigma_-,
\label{ides}
\end{equation}
{\it i.e.} the $U$ Lax operator   of the NLSE (\ref{U}).

In addition, the Lax operator $R_t$ can be obtained, by simply identifying it 
with the $V_0$ Lax operator  of the NLSE (\ref{fadV0}), that is, 
\begin{equation*}
R_t \ = \ i\lambda |q|^2 \sigma_3 - i\sqrt{\lambda} \left( \partial_s \bar q \sigma_+ - \partial_s q \sigma_-\right).
\end{equation*}
Thus, the integrability condition (\ref{Fts2}) is  satisfied for $q$ being a solution of the NLSE (\ref{nlsef}).
Therefore, an {\it embedding} of  
the NLSE Lax pair in the Lax pair of the spinor Frenet equation is derived.  In particular,
the NLSE determines the time evolution of the string.

We conclude, by pointing  out that    the complex spectral parameter $\xi$ can be interpreted as a parameter for a family of loops in the group manifold,
determined by 
\[
g(s) \mapsto g(\xi,s)
\]
with 
\begin{equation*}
R(\xi,s)= g^{-1}(\xi,s) \, d g(\xi,s).
\end{equation*}
The real part of  $\xi$ 
parameterizes  a gauge transformation {\it i.e.} a rotation
of the parallel transport frame, that proceeds
linearly in the arc-length parameter $s$. 
The imaginary part of $\xi$ corresponds to a Weyl transformation (\ref{infi}) and  (\ref{fe}) {\it i.e.}  a rescaling of the string that similarly proceeds linearly in the arc-length parameter.

%
%
%
%
%
%

\section{String Hamiltonians}

Once a Lax pair of an integrable model is known, the conserved charges can be constructed. 
The procedure is standard, it utilizes an expansion in the spectral parameter
and has been studied in detail in \cite{faddetak}.
In particular, an infinite number of string Hamiltonians can be constructed from the Lax 
pair ($R_s, R_t$) of the spinor Frenet equation (obtained from   the NLSE
Lax pair). 

More generally, {\it any} SU(2) Lax pair of a one dimensional integrable model can be utilized to construct a time evolution of a string. To do so, equations (\ref{Fts2}) and (\ref{Rlax1})   can be  used to identify ($R_s$, $R_t$) in  the spinor Frenet equations (\ref{Psis}) and (\ref{Psit}) in terms of the Lax pair of the given integrable model, similar to the NLSE case.

In what follows it is shown how string Hamiltonians in the NLSE hierarchy can be obtained alternatively, using a formalism of projection operators; introduced originally in the  integrable $\mathbb C \mathbb P^{\mathrm N}$ 
models. 
Note that the spinorial  approach to strings engages locally the structure  of $\mathbb S^2 \times 
\mathbb S^1$.  Since $\mathbb S^2 \simeq \mathbb C \mathbb P^1$,  
the connection between the projection operator 
formalism of the $\mathbb C \mathbb P^1$ model and the spinor Frenet equation can be explored.

%
%
%
%
%
%

\subsection{Projection operators}

Let us start with the component representations (\ref{psi1}) and (\ref{C}). Using  the matrix $g \in $ SU(2) of (\ref{g}) one can easily show that 
\[
\psi \ = \ g \left( \begin{matrix} 1 \\ 0 \end{matrix} \right), \ \ \ \ \ \ \ \ \ \ \ \ \ \bar \psi \ = \ g \left( \begin{matrix} 0 \\ 1 \end{matrix} \right).
\]
Then, the   projection operator can be defined as 
\begin{equation*}
\mathbb P \ = \ |\psi\!\!> <\!\!\psi |  \ \equiv \ \psi \!\otimes \! \psi^\dagger \ = \ g \left( \begin{matrix} 1 & 0 \\ 0 & 0 
\end{matrix} \right) g^{-1}
\end{equation*}
which has been utilized widely, in the context of $\mathbb C \mathbb P^1$ (more generally $\mathbb C \mathbb P^N$)  model \cite{berg}-\cite{wojtek}.
 
In addition,   introduce the complemental projection operator
\begin{equation*}
 \bar{\mathbb P}\ = \  |\bar\psi\!\!> \!
\otimes \!<\!\!\bar\psi | \ \equiv \ \bar\psi \!\otimes \! \bar{\psi}^\dagger \ = \ g \left( \begin{matrix} 0 & 0 \\ 0 & 1 
\end{matrix} \right) g^{-1}.
\end{equation*}
The projection operators satisfy the following identities
\begin{eqnarray*}
\mathbb P ^2 \ & = &  \ \mathbb P, \\
 \bar{ \mathbb P} ^2 \ & = & \ \bar{\mathbb P},\\
\mathbb P \, \bar{ \mathbb P} \ & = & \ \bar{\mathbb P}
\, \mathbb P \ = \ 0, \\
\mathbb P +  \bar{ \mathbb P } \ & = & \  \mathbb I.
\end{eqnarray*}
Finally,  introduce the nilpotent operators
\begin{eqnarray*}
\mathbb Q & = &  |\psi\!\!>\! \otimes \!<\!\!\bar\psi|  \ \equiv \ \psi \!\otimes \! {\bar \psi}^\dagger
\ = \ g \left( \begin{matrix} 0 & 1 \\ 0 & 0 
\end{matrix} \right) g^{-1}
\\
 \bar{\mathbb Q} &  = & |\bar \psi\!\!> \!\otimes \!<\!\!\psi | \ \equiv \ 
 \bar\psi \!\otimes \! {\psi}^\dagger \ = \ g \left( \begin{matrix} 0 & 0 \\ 1 & 0 
\end{matrix} \right) g^{-1}
\end{eqnarray*}
where
\[
{\mathbb Q}^2 \ = \ \bar{\mathbb Q}^2 \ = \ 0.
\]
The $\mathbb Q$ and  $ \bar{\mathbb Q}$ exchange the spinors since  
\begin{eqnarray*}
\mathbb Q \, \bar \psi \ & = & \ \psi, \\  \mathbb Q \, \psi \ & = & \  0,\\
 \bar{\mathbb Q}\, \psi \ & = & \ \bar \psi, \\   \bar{\mathbb Q}\,  \bar\psi \  &= & \ 0.
\end{eqnarray*}
Also the projection and nilpotent operators are related through the relations
\begin{eqnarray*}
\bar{\mathbb Q} \, \mathbb Q \ & = & \ \bar{ \mathbb P},  \\
\mathbb Q \, \bar{ \mathbb Q}  \ & = & \ {\mathbb P}, \\
\mathbb Q \, \mathbb P \ & = & \ \bar{ \mathbb P} \, \mathbb Q  \ = \ 0,  \\
\mathbb Q \, \bar{\mathbb P} \ & = & \ 
\mathbb P \, \mathbb Q \ = \  \mathbb Q. 
\end{eqnarray*}

In terms of the  \underline{su}(2) Lie algebra  generators (\ref{su2}), one can set
\begin{eqnarray}
&&\mathbb P -  \bar{ \mathbb P } \ \simeq  \ \hat{\mathbf t} 
 \label{Pandt}
 \\
&&\mathbb Q \ \simeq \ \hat{\mathbf e}_+  
\label{Qand+}
\\
&& \bar{\mathbb Q} \ \simeq \ \hat{\mathbf e}_- 
\label{Qand-}
\end{eqnarray}
and note that the U(1) transformations  (\ref{u11}) and  (\ref{gh}) leads to 
\begin{eqnarray}
\mathbb Q & \ \ \buildrel{\varphi}\over{\longrightarrow} & \ e^{2i\varphi} \mathbb Q 
\label{U1QA}
\\
\bar{\mathbb Q} & \ \ \buildrel{\varphi}\over{\longrightarrow}  & \ e^{-2i\varphi} \bar{ \mathbb Q}
\label{U1QB}
\end{eqnarray}
while the projection operators  $\mathbb P$ and $\bar{\mathbb P}$ remain intact under the U(1) transformation. 

Then the following relations between the derivatives of these operators and the 
left Maurer-Cartan form (\ref{L}) exist
\begin{eqnarray}
 (\mathbb P- \bar{\mathbb P})_s &= & [ \, L \, , \, \mathbb P- \bar{\mathbb P} \, ]
\ = \ \kappa_c  \hat{\mathbf e}_- +  \kappa_c^\star  \hat{\mathbf e}_+, 
\label{darP}
\\
\mathbb Q_s &= &  [ \, L \, , \, \mathbb Q \, ] \ = \  - i \tau_r \hat{\mathbf e}_+  - \frac{\kappa_c}{2} \hat{\mathbf t}, 
\\
 \bar{\mathbb Q}_s & = &  [ \, L \, , \, \bar{\mathbb Q} \, ]  \ = \   i \tau_r \hat{\mathbf e}_-  - \frac{\kappa_c^\star}{2} \hat{\mathbf t}
 \label{darQ2}
\end{eqnarray}
which, due to  (\ref{Pandt})-(\ref{Qand-}), can be identified as  the spinor Frenet equation,
in the dreibein Darboux format (\ref{contso3}).  
By defining the  $L$-covariant derivative
\[
D_L \ = \ \partial_s - [ \, L \, , ~ ~]
\]
one can re-write the spinor Frenet equation compactly as
\[
\begin{matrix}
D_L (\mathbb P- \bar{\mathbb P}) & = &  0 \\ \\
 D_L  \mathbb Q & = & 0 \\ \\
 D_L \bar{\mathbb Q}& = & 0.
 \end{matrix}
\]
Furthermore, since ($\hat{\mathbf e}_\pm , \hat{\mathbf t}$) span the \underline{su}(2) Lie algebra, we (\ref{darP})-(\ref{darQ2}) can be reverted so that $L$ can be represented in terms of the operators (\ref{Pandt})-(\ref{Qand-}) as
\begin{eqnarray}
&&\frac{i}{4} [ \, \mathbb P- \bar{\mathbb P} \, , \, (\mathbb P- \bar{\mathbb P})_s ] + 
\frac{i}{2}  [ \, \mathbb Q \, , \, \bar{\mathbb Q}_s \, ] + \frac{i}{2}  [ \, \bar{\mathbb Q} \, , \, {\mathbb Q}_s \, ] 
\label{laxp1}
\\
& & =  - \tau_r \, (\mathbb P- \bar{\mathbb P}) +i \kappa_c^\star  \mathbb Q   - i \kappa_c
\bar{\mathbb Q}\nonumber\\
& &
\equiv \ - \tau_r \, \hat{\mathbf t}  +i \kappa_c^\star \hat{\mathbf e}_+   - i \kappa_c
\hat{\mathbf e}_-
\nonumber\\
& &\equiv  \mathfrak a \, \hat{\mathbf t} + \mathfrak b_+ \hat{\mathbf e}_+  + \mathfrak b_-  \hat{\mathbf e}_- \nonumber\\
& & \equiv - 2i L.
\label{PQs}
\end{eqnarray}
Since the left Maurer-Cartan form $L$ relates to (\ref{ides}) by a combination of the gauge transformations (\ref{L}), (\ref{3abr2}) and   (\ref{partran2}) one of  the two Lax pair operators  of NLSE is obtained.

The other Lax pair operator of NLSE is obtained as follows: First observe that
\begin{equation}
\mathbb P_{ss} \, = \, \frac{1}{2}\left[  {\bar D}_s \kappa^\star_c {\mathbb Q} +   D_s \kappa_c \bar{ \mathbb Q}
-  |\kappa_c|^2 (  \mathbb P- \bar{\mathbb P} )\right]
\label{QPs}
\end{equation}
where  the covariant derivative is given by 
\[
D_s = \partial_s + i \tau_r.
\]
By  introducing  the U(1) gauge transformation (\ref{3abr}),  (\ref{U1QA}), (\ref{U1QB}), the operator (\ref{QPs}) is gauge invariant. Thus, by choosing 
\[
\varphi \ = \ - \frac{1}{2} \int^s \! \tau_r (s') ds'
\]
{\it i.e.}  using the parallel transport frame, one gets the explicit representation
\begin{equation}
\mathbb P_{ss} \, = \, \frac{1}{2} \left[  \partial_s \kappa^\star_c  \hat {\mathbf e}_+  +  \partial_s \kappa_c
\hat{\mathbf e}_-
-  |\kappa_c|^2 \hat{\mathbf t}\right].
\label{QPs2}
\end{equation}
Similar to the case of  (\ref{PQs}), recalling  (\ref{kappaide}),  the left Maurer-Cartan realization of the second NLSE Lax pair operator is obtained. 
Explicitly,  by applying   the  U(1) transformation (\ref{3abr2})  to (\ref{QPs2}) and using
the identification (\ref{kappaide}) leads to
\[
\mathbb P_{ss} \  \buildrel h \over \longrightarrow \ \mathbb P^h_{ss}  
\]
where 
\begin{eqnarray*}
\mathbb P^h_{ss} &=&  \ - \frac{i}{2} \sqrt{\lambda}\partial_s\bar q \, \sigma_+  + 
\frac{i}{2} \sqrt{\lambda}\partial_s q \, \sigma_- - \frac{1}{2} \lambda |q|^2 \sigma^3
\\
& \equiv& \ \frac{i}{2} V_0
\end{eqnarray*}
{\it i.e.} equation (\ref{fadV0}).

\subsection{Hamiltonians}

From the previous subsection, we conclude that the NLSE Lax operators and  
the conserved quantities in the NLSE hierarchy can be constructed in terms of the two operators (\ref{PQs}) and (\ref{QPs2}). However, in generic Frenet frame, 
the Lax pair constitutes the two operators (\ref{laxp1}) and (\ref{QPs}).
Thus, the conserved charges can be expected to be combinations of the 
projection operators $\mathbb P$, $\bar{\mathbb P}$;
the exchange operators $\mathbb Q$, $\bar{\mathbb Q}$; and their (covariant) derivatives. 
Next, we proceed to elaborate on the relations, beyond the Lax pair.

An example of a familiar Hamiltonian density is  the ``number operator" in the NLSE hierarchy {\it i.e.} 
Heisenberg spin chain Hamiltonian
\begin{equation}
\mathcal H_2 \ = \ \mbox{tr} \left\{ \mathbb P_s \mathbb P_s \right\} \ = \  \frac{1}{2}|\kappa_c|^2 \ = \ \frac{1}{2}
\partial_s \mathbf t \cdot \partial_s \mathbf t.
\label{ham1}
\end{equation}
Using (\ref{frenet1}) and (\ref{taur}), this can be further presented as
\[
{\mathcal H_1} \ = \ \frac{1}{2}\left |(\partial_ s + \frac{i}{2} \tau_r)    \psi ) \right |^2 \ = \ 
\frac{1}{2} \left|\left(\partial_ s - <\!\psi^\dagger, \partial_s \psi\!> \right) \psi \right|^2
\]
{\it i.e.} the Hamiltonian of the $\mathbb C \mathbb P^1$ model.

Another familiar example is  the NLSE Hamiltonian given by
\begin{equation}
{\mathcal H_3} \ = \ \mbox{ tr} \left\{ \mathbb P_{ss} \mathbb P_{ss}\right \} = \frac{1}{2} \left|  D_s \kappa_c \right|^2 +  \left|\kappa_c\right|^4.
\label{ham2}
\end{equation}

The following manifestly U(1) gauge invariant 
``Lax" pair operators have been 
considered in the context of the $\mathbb C \mathbb P^1$ model \cite{wojtek}:
\begin{eqnarray}
\mathbb L(\lambda)  = 2\,  \frac{ [\partial_s \mathbb P \, , \, \mathbb P] }{1+\lambda} \ \equiv  \ 
\frac{\mathbb L}{1+\lambda} 
\label{lax1}
\\
\bar{\mathbb L}(\lambda)  = 2 \, \frac{ [\partial_s \bar{\mathbb P} \, , \, \bar{\mathbb P}] }{1+\lambda} \ \equiv  \ 
\frac{\bar{\mathbb L}}{1+\lambda}.
\label{lax2}
\end{eqnarray}
In terms of the exchange operators 
\[
{\mathbb L }\ = \ \kappa_c  \bar{\mathbb Q}  \, - \, \kappa_c^\star   \mathbb Q 
\]
and using the properties of the  projection and exchange operators, it can be shown that
\begin{eqnarray*}
 \frac{1}{2}\,  \mathbb L^{2n}   &=&  (-1)^n  |\kappa_c |^{2n}  \, \mathbb I \\
 \mbox{tr} \left( \mathbb L^{2n+1} \right)  &= & 0.
\end{eqnarray*}
Thus not all conserved charges in the NLSE hierarchy can be obtained 
from (\ref{lax1}) and (\ref{lax2}). In particular, the NLSE Hamiltonian (\ref{ham2})  can not be
presented in terms of polynomials of these operators which means that derivatives of $\mathbb L$ also need to be introduced: 
\[
{\mathbb L_s} \ = \ D_s \kappa_c \, \bar{\mathbb Q} - (D_s \kappa_c )^\star \mathbb Q.
\]
Then
\begin{eqnarray*}
\frac{1}{2} \, \mathbb L_s^{2n} &=&  (-1)^n |D_s \kappa_c |^{2n} \, \mathbb I \\
\mbox{ tr} \left( \mathbb L_s^{2n+1} \right) & = & 0.
\end{eqnarray*}
In this case, using combinations of $\mathbb L$ and $\mathbb L_s$ various U(1) gauge invariant  and conserved densities can be constructed. 

For example,  the NLSE Hamiltonian (\ref{ham2})  can also be presented  as
\[
{\mathcal H_3} \,{ \mathbb I  }\ = \ - \frac{1}{2} \mathbb L_s^2 + \gamma  \, \mathbb L^4
\]
where $\gamma$ is a parameter while the number operator (\ref{ham1}) becomes
\[
\mathcal H_1\, \mathbb I   \ = \ - \frac{1}{4} \mathbb L^2.
\]
Furthermore,  in analogy with  (\ref{lax1})  by introducing  the  U(1) gauge invariant operator
\[
\mathbb T = [ \, \mathbb L_s \, , \, \mathbb L \, ]  
\]
the gauge invariant conserved momentum density of the NLSE 
hierarchy
\[
\mathbb T \, = \, 8 i\,  {\mathcal H_2} \left(\bar{\mathbb P} - \mathbb P\right)
\]
is derived. 
Thus, the  conserved momentum density can be presented as follows
\begin{eqnarray*}
\mathcal H_2 \, \mathbb I  & = & \frac{i}{8}\left\{ {\mathbb P} \mathbb T - \mathbb T \bar{\mathbb P}\right\}\nonumber \\
& = & - \frac{1}{4} \left\{ \tau_r |\kappa_c|^2 
- \frac{1}{2} ( \kappa_c \partial_s \kappa_c^\star -  \partial_s \kappa_c\kappa_c ^\star)\right\}.
\end{eqnarray*}
Recall that  the  U(1) invariant  ``super-current" that appears 
\begin{equation*}
J \ = \  \tau_r - \frac{1}{2} \, \frac{ \kappa_c 
\partial_s \kappa_c^\star -  \partial_s \kappa_c\kappa_c ^\star}{|\kappa_c |^2}  
\end{equation*}
 is akin to the super-current  introduced in (\ref{superj}).

The above construction  does not exhaust the full set of conserved charges of the NLSE hierarchy.
In particular, for the conserved helicity (which can not be derived from the standard NLSE Lax pair) one concludes that
\begin{eqnarray*}
\mathbb Q_s \mathbb Q &=&  - \frac{\kappa_c}{2} \mathbb Q\\
\mathbb P_s \mathbb P &=&   \frac{\kappa_c}{2} \bar{\mathbb Q},
\end{eqnarray*}
leading  to the equation
\[
[\mathbb P, \mathbb Q_s] \ = \ -i\tau_r  \mathbb Q.
\]
That way  the helicity density can be derived since 
\begin{equation}
\mathcal H_{-2} \ = \ 
\left\{ \bar{\mathbb Q} \, , \mathbb Q_s \right \} = -i\tau_r \, \mathbb I.
\label{ham3}
\end{equation}
Its integral  is  a conserved quantity  in the NLSE 
hierarchy, and also  invariant under the U(1) frame rotation (\ref{u1tau}) provided that
\[
\varphi (0) = \varphi(L) = 0.
\]
It is notable that a gauge transformation such as (\ref{partran2}), does not preserve
helicity. 

Finally, let us point out that the length of the string 
\[
L = \int_0^L \!\!\! ds\,  \sqrt{ \dot {\mathbf x}\cdot \dot {\mathbf x}} \ \equiv \  \int_0^L \!\!\! ds\,   \mathcal H_{-1}
\] 
also appears as a conserved charge in the NLSE hierarchy, albeit it can not be derived from the NLSE Lax pairs. 
Its density can be identified {\it e.g.} with 
\[
\mathcal H_{-1} \simeq \mathbb P + \bar{\mathbb P}.
\]
Note that $ \mathcal H_{-1}$ relates intimately to the Nambu action, and thus to the Polyakov action of a relativistic string. More generally, a combination of
$ \mathcal H_{-1}$ and $ \mathcal H_{1}$ introduced in (\ref{ham1}) constitutes the essence of 
Polyakov's rigid string action \cite{poly}.

\section{Decomposing Lax pair}

In what follows let us  consider a generic Riemann
surface in $\mathcal S \in \mathbb R^3$. In the sequel, ($u,v$) denote generic coordinates on the Riemann 
surface.  
Two components of the \underline{su}(2) Lie algebra valued
right Maurer-Cartan form the following currents
\[
R_u \ = \ g^{-1} \partial_u  g,  \ \ \ \ \ \ \ \ \ R_{v} \ = \ g^{-1} \partial_{v} g
\]
which satisfy the   Lax pair. However,  for a  flat connection the zero-curvature condition (\ref{Fts2}) becomes an identity, and there is no equation left to be satisfied.  In order to construct a  Lax pair  yielding an integrable system, a Maurer-Cartan form needs to be deformed. 

\subsection{Abelian Higgs Model}

Let us start with the following  decomposition of a generic two-dimensional SU(2) Yang-Mills
connection  \cite{fadde1}
\begin{equation}
A_{\alpha}^a = C_{\alpha} n^a + \epsilon^{abc} \partial_{\alpha} n^b n^c +  \rho_1
\partial_{\alpha} n^a + \rho_2 \epsilon^{abc}
\partial_{\alpha} n^b n^c
\label{fadec1}
\end{equation}
where  $\alpha = 1,2$  correspond to   $u,v$;  $C$ is a two-component one-form; and
\[
\rho = \rho_1 + i \rho_2
\] 
is a complex field. Note that  for  $\rho = 0$ and $C$ identified with  (\ref{3aco}),
the decomposition (\ref{fadec1}) becomes a left Maurer-Cartan form, {\it e.g.} like  (\ref{Lte}).  In particular, a nontrivial multiplet ($C_\mu , \rho$) specifies a deformation of the Maurer-Cartan form.

We first argue that (\ref{fadec1}) is a complete decomposition of the two dimensional SU(2) connection.
To see that simply observe that in the {\it l.h.s} of (\ref{fadec1}) there are six real valued 
field degrees of freedom.  On the other hand, in the {\it r.h.s.} 
decomposition
there are: two real field degrees of freedom in the complex field $\rho$; 
two in the one-form $C$; and  two in the unit vector $n^a$.  Thus the number of field degrees of freedom match. 
A detailed proof of the completeness is presented in \cite{oma3d}.

Note that   under the gauge transformation around the direction of $\mathbf n$
\begin{equation*}
h \ = \ \exp\{ i \theta \, \mathbf n \cdot \mathbf  \sigma\}
\end{equation*}
the functional form (\ref{fadec1}) remains intact when redefining
\begin{eqnarray*}
C_\alpha & \buildrel{h}  \over{\longrightarrow}  &  C_\alpha +  2 \partial_{\alpha} \theta
\\
\rho &  \buildrel{h}  \over{\longrightarrow} &  e^{ -2i \theta} \rho.
\end{eqnarray*}
This follows due to the property of the decomposition under a gauge transformation. 
Thus ($C, \rho$) comprises an  Abelian Higgs multiplet.

In accordance with  (\ref{gmat1}) let us introduce a $g\in$ SU(2) matrix defined by
\begin{equation}
\hat {\mathbf n} \ = \ g \, \sigma^3 \! g^{-1}
\label{halpha}
\end{equation}
so that  the decomposition (\ref{fadec1})  can be written as
\begin{equation}
A =  g \left( C\sigma^3 + i R^{diag} + \rho_1 [ R,\sigma^3] - i \rho_2 R^{off}\right) g^{-1} - i L.
\label{gaudec}
\end{equation}
Here $R^{diag}$ is the diagonal part,  $R^{off}$ is the  off-diagonal part of the right Maurer-Cartan form $R$ and, $L$ is the left Maurer-Cartan form of $g$. Thus the decomposed  connection (\ref{gaudec}) is gauge equivalent to
\begin{equation*}
B = C \sigma^3 + i R^{diag} + \rho_1 [ R,\sigma^3] - i \rho_2 R^{off}. 
\end{equation*}
We remark  that $g$ is not unique, there is the U(1) latitude (\ref{gh}) and  (\ref{3nro})
which leaves (\ref{halpha}) invariant under the following right multiplication
\[
g \ \to \ g\, e^{i\theta \sigma^3}.  
\]

Note that, using the local coordinate parametrization 
\begin{equation}
\mathbf n \ = \ \left( \begin{matrix} \cos \varphi \sin\vartheta \\ \sin\varphi \sin\vartheta \\ \cos\vartheta 
\end{matrix} \right)
\label{nvecp}
\end{equation}
and introducing the notation (\ref{calb}), the off-diagonal components of $B$ take  the form
\begin{equation}
B^\pm \ = \ \left(\rho_1 \pm i \rho_2\right) \left(d\vartheta \mp i \sin\vartheta d\varphi\right) \ 
\simeq \ \left(\rho_1 \pm i \rho_2\right) \mathfrak b_\pm.
\label{B+-}
\end{equation}
 Thus
(\ref{fadec1}) is gauge equivalent to the  deformation of the right Cartan form
\[
R \to  A
\]
where
\begin{eqnarray*}
R  &= & \mathfrak a \sigma^3 + \mathfrak b_+ \sigma^+ + \mathfrak b_- \sigma^-\\
 A & = & C \sigma^3 + \rho\, \mathfrak b_+\sigma^+ + \rho^\star \mathfrak b_-\sigma^-
\equiv  \ C \sigma^3 + B_+\sigma^+ + B_-\sigma^-.
\end{eqnarray*}
Then the action of the U(1) rotation on its components is of the form
\begin{equation}
A \ \buildrel{e^{i\theta \sigma^3}}\over{\longrightarrow} \ 
(C + 2 d \theta)  \sigma^3 + e^{2i\theta} B_+\sigma^+ + e^{-2i\theta} 
B_-\sigma^-.
\label{gaugeA}
\end{equation}

By substituting  (\ref{fadec1}) in the curvature two-form
\[
F^a_{\alpha\beta}  = \partial_\alpha A^a_{\beta} - \partial_{\beta} A^a_\alpha+  \epsilon^{abc} A^b_\alpha A^c_{\beta}
\]
one gets
\begin{eqnarray}
F^a_{\alpha\beta} &=& n^a \left( G_{\alpha\beta} - \left[ 1 - 
\rho\bar\rho\right] H_{\alpha\beta} \right) \nonumber\\
&+& \left\{ (\partial_\alpha \rho_1 - C_\alpha \rho_2) \partial_{\beta} n^a
- (\partial_{\beta} \rho_1 - C_{\beta} \rho_2) \partial_{\alpha} n^a \right\}\nonumber\\
&+& \epsilon^{abc}n^b
  \left\{  (\partial_\alpha \rho_2 + C_\alpha \rho_1) \partial_{\beta} n^c - 
(\partial_{\beta} \rho_2 + C_{\beta} \rho_1) \partial_{\alpha} n^c\right\}
\label{Frho}
\nonumber\\
\end{eqnarray}
where
\[
\begin{matrix}
G_{\alpha\beta} & = & \partial_\alpha C_{\beta} - \partial_{\beta}C_{\alpha} \\ \\
H_{\alpha\beta} & = & \epsilon_{abc} n^a  \partial_\alpha  n^b  \partial_{\beta} n^c 
\end{matrix}
\]
and 
\begin{eqnarray*}
D_\alpha \rho  &=&  (\partial_\alpha + i C_\alpha) \rho\nonumber \\
&= &  (\partial_\alpha \rho_1 - C_\alpha \rho_2) + i (\partial_\alpha \rho_2 + C_\alpha \rho_1). 
\end{eqnarray*}

But the curvature  (\ref{Frho})  of  the Lax pair of an integrable system must vanish
\[
F^a_{\alpha\beta} = 0
\] 
which leads to the system of equations
\begin{eqnarray}
G_{\alpha\beta} - \left( 1 - \bar \rho \rho\right) H_{\alpha\beta} &= & 0 \nonumber\\ 
D_\alpha \rho  &= & 0.
\label{bogo}
\end{eqnarray}
The above equations have the form of  the Bogomolny equations \cite{bogomolny}-\cite{pauls}
for the energy function of the Abelian Higgs Model on a Riemann surface
with metric 
\[
g_{\alpha\beta}=\int_{\mathcal S} \!  \!
\sqrt{g}\, dudv \left\{ \frac{1}{4} G^2_{\alpha\beta} + 
|D_\alpha \rho|^2 + (1-|\rho|^2)^2 \right\}.
\]
Vortex solutions to these equations in the background of a given Riemann surface
have been studied extensively  in  \cite{pauls}-\cite{nick2}, to which we refer to explicit constructions. 

Here it suffices to make the following remarks:
Assume that $\mathbf x(u,v)$ describes the points on the surface of $\mathcal S \in \mathbb R^3$, and that the
two tangent vector fields $\partial_u \mathbf x$ and $\partial_v \mathbf x$ are linearly independent. 
Let us  identify $\mathbf n$ as the normal vector of $\mathcal S$
\[
\mathbf n \ = \ \frac{  \mathbf x_u \times \mathbf x_v }{||  \mathbf x_u \times \mathbf x_v ||},
\]
which  defines the Gau\ss ~ map 
\begin{equation}
\mathbf n: \ \ \ \mathcal S \mapsto \mathbb S^2.
\label{gausmap}
\end{equation}
Then
\begin{equation}
\partial_u \mathbf n \times \partial_v \mathbf n \ = \ \frac{1}{2} R
 \, \partial_u \mathbf x \times \partial_v \mathbf x \ \equiv \ K 
  \, \partial_u \mathbf x \times \partial_v \mathbf x
\label{RK}
\end{equation}
where $ R $ is the scalar curvature which is twice the Gau\ss ian curvature $K$ of $\mathcal S$.

Next let us   introduce the zweibein field ${e^i}_\alpha$,
\begin{equation*}
\delta_{ij}{e^i}_\alpha{e^j}_\beta \ = \ g_{\alpha\beta} \ = \ \delta_{\mu\nu}
\partial_\alpha x^\mu \partial_\beta x^\nu \ \ \ \  \ \ (\mu,\nu = 1,2,3)
\end{equation*}
where $g_{\alpha\beta}$ is the induced metric.
For  the spin connection $\omega^i_{\alpha j}$ first  define
\[
\omega_{\alpha } =   \frac{1}{\sqrt{g}} \delta_{ij} {e^i}_\alpha \epsilon^{\beta\gamma }  \partial_\gamma {e^j}_\beta  
\] 
and then set 
\[
\omega_{\alpha j}^i \ = \ \omega_\alpha {\epsilon^i}_j.
\]
Note that then the Gau\ss~map,  the spin connection and the scalar curvature of $\mathcal S$ are related since
\[
\frac{1}{2} \epsilon_{abc} \epsilon^{\alpha\beta}n^a \partial_\alpha n^b \partial_\beta n^c \ = \ 
 \epsilon^{\alpha\beta} \partial_\alpha \omega_\beta  \ = \ \frac{1}{2} \sqrt{g}\,  R. 
  \]

Now by considering  the combined Weyl transformation and SO(2) rotation such that 
\begin{equation}
{e^i}_\alpha \ \mapsto \ e^{\frac{1}{2} \phi}{e^j}_\alpha \left( {\delta^i}_j \cos \frac{\theta}{2} - {\epsilon^i}_j \sin
\frac{\theta}{2}\right)
\label{weylso2}
\end{equation}
the metric tensor, the spin connection and the scalar curvature of $\mathcal S$ transform to 
\begin{eqnarray*}
g_{\alpha\beta} & \mapsto &  e^\phi g_{\alpha\beta}\\
\omega_\alpha & \mapsto &  \omega_\alpha + \sqrt{g}\, {\epsilon^\beta}_\alpha \partial_\beta \phi +  \partial_\alpha \theta\\
R  & \mapsto & e^{-\phi}\!  \left( R + \Delta_g \phi \right) 
\end{eqnarray*}
where
\[
\Delta_g = - \frac{1}{\sqrt{g}} \partial_\alpha ( \sqrt{g} g^{\alpha\beta} \partial_\beta)
\]
is the Laplacian on $\mathcal S$.

Finally, let us   write the first equation (\ref{bogo}) as 
\begin{equation}
G \ = \ \frac{1}{2}  \left(1-\rho\bar\rho \right) \sqrt{g} R 
\label{bogo1}
\end{equation}
and assume that the surface $\mathcal S$ is closed. By integrating  (\ref{bogo1}) over the surface one gets
\begin{eqnarray}
{\mathrm C \mathrm h_1} [C] & \buildrel{def}\over{=} & \frac{1}{2\pi} \int_{\mathcal S} G
\nonumber\\
& = &  \frac{1}{4\pi} \int_{\mathcal S} \sqrt{g} R - 
\frac{1}{4\pi} \int_{\mathcal S}\rho\bar\rho \sqrt{g} R \nonumber \\
&\buildrel{def}\over{=} & \mathcal X - \frac{1}{4\pi} \int_{\mathcal S}\rho\bar\rho \sqrt{g} R
\label{Chernchar}
\end{eqnarray}
where ${\mathrm C \mathrm h_1}[C]$ is the first Chern character of $G$ and $\mathcal X$ is the Euler character of $\mathcal S$.
Note that although for a  compact surface with no boundary each is an  integer,  in the case of {\it e.g.} hyperbolic manifolds this  does not need to be the case.

 The effect of the Weyl transformation (\ref{weylso2}) in the relation
\[
\frac{1}{4\pi} \int_{\mathcal S}\rho\bar\rho \sqrt{g} R \ = \ \mathcal X - {\mathrm C \mathrm h_1} 
\]
leaves intact the  two quantities in the {\it r.h.s.} while   the {\it l.h.s.} quantity becomes
\[
- \frac{1}{4\pi} \int_{\mathcal S} \Delta_g (\rho\bar \rho) \phi
\]
which, for general $\phi$, vanishes only if $\rho \bar \rho$ is harmonic. 

On the other hand, in terms of the complex coordinate $z=u+iv$ the second equation (\ref{bogo}) becomes
\begin{equation*}
D_z \rho \ = \ (\partial_z + i C_z) \rho \ = \ 0 
\end{equation*}
implying that the compination
\[
 \rho \, e^{- i \int\limits^z C_z }
\]
is (anti-)holomorphic. 

In general, the product of two holomorphic functions is {\it not} harmonic. However,
examples of  non-trivial solutions can  be obtained, with constant $|\rho|\not=1 $. {\it E.g.} for $\mathcal S$ being a punctured Riemann surface; or for  $\mathcal S$ being  a hyperbolic Riemann surface;  or, more generally, for $\mathcal S$ being  a Riemann surface that is not simply connected. 

In such cases, there are generically loops for which the line integral
\[
\oint d{\vec{l}} \cdot \vec C
\]
does not need to vanish, and the solution to the first equation (\ref{bogo}) is
\[
C = (1-|\rho|^2)   d^{-1} H.
\]
For example, if $\rho$ vanishes on $\mathcal S$ the one-form $C$ is like the Dirac monopole field and (\ref{fadec1}) coincides with (\ref{Lte}). 

\subsection{Surfaces in $\mathbb R^3$}

Next the relations between the two dimensional decomposed Yang-Mills theory and the Gau\ss-Godazzi equations is considered.

Let  $\mathbf x(u,v)$ describe the points of a Riemann surface $\mathcal S$  in $ \mathbb R^3$, in terms of the local coordinates  ($u,v$) on $\mathcal S$.  Then  the Gau\ss-Codazzi and   the Weingarten equation are given by 
\begin{eqnarray}
\mathbf x_{\alpha\beta} \ = \ \Gamma^\gamma_{\alpha\beta} \mathbf x_\gamma + h_{\alpha\beta} \mathbf n
\label{gaussc}
\\
\mathbf n_\alpha = - h_{\alpha\beta}g^{\beta\gamma}\mathbf x_\gamma \ \equiv - {h_{\alpha}}^\beta\mathbf x_\beta,
\label{weing}
\end{eqnarray}
respectively.
Here \[
\Gamma^\alpha_{\beta\gamma} \ = \ \Gamma^\alpha_{\gamma\beta} \ = \ 
\frac{1}{2} g^{\alpha\delta}(g_{\delta\beta,\gamma} +
g_{\delta\gamma,\beta} - g_{\beta\gamma,\delta} )
\]
are the Christoffel symbols, $\mathbf n$ is the normal vector to the surface and
$g_{\alpha\beta}$ is taken to be the (induced) metric
\[
g_{\alpha\beta} \ = \ \partial_\alpha \mathbf x \cdot \partial_\beta \mathbf x.
\]
Finally,  $h_{\alpha\beta}$ are the components of the
second fundamental form
\[
h_{\alpha\beta} \ = \ h_{\beta\alpha} \ = \ 
\left( \begin{matrix} \mathbf x_{uu}\!\cdot \!\mathbf n & \mathbf x_{uv}\!\cdot \!\mathbf n \\
\mathbf x_{vu}\!\cdot \!\mathbf n & \mathbf x_{vv}\!\cdot \!\mathbf n\end{matrix} \right).
\]
Note that equations (\ref{gaussc}) and (\ref{weing}) can be combined  into the matrix equation
\begin{equation}
\partial_\alpha  \left( \begin{matrix} \mathbf x  \\ \mathbf x_u \\ \mathbf x_v \\ \mathbf n \end{matrix} \right)
\ = \ \left( \begin{matrix} 0 & \delta_{\alpha u}  & \delta_{\alpha v} & 0 \\
0 & \Gamma_{\alpha u}^u & \Gamma_{\alpha u}^v & h_{u \alpha } \\
0 & \Gamma_{\alpha v}^u & \Gamma_{\alpha v}^v & h_{ v \alpha } \\
0 & - {h_\alpha}^u  &   - {h_\alpha}^v  & 0 \end{matrix} \right)
   \left( \begin{matrix} \mathbf x \\\mathbf x_u \\ \mathbf x_v \\ \mathbf n \end{matrix} \right).
\label{surf1}
\end{equation}

In what follows, let us assume that at each point $\mathbf x(u,v)$ of the surface we have an orthonormal frame ($\mathbf e_1, \mathbf e_2, \mathbf n$) in $\mathbb R^3$, where $\mathbf n$ is  the Gau\ss~map ({\it i.e.} a unit normal vector). 

For example, by  choosing the  coordinates ($u,v$) so that the 
vectors $\mathbf e_1$ and $\mathbf e_2$ point into the principal directions
\begin{eqnarray}
\mathbf e_1 & = & \frac{\mathbf x_u}{|| \mathbf x_u||}
\label{e1e21} 
\\
\mathbf e_2   & = &  \frac{\mathbf x_v}{|| \mathbf x_v||} 
\label{e1e22} 
\\
\mathbf n   & = &  \ \mathbf e_1 \times \mathbf e_2
\label{e1e2n}
\end{eqnarray}
the corresponding integral  curves are the lines of curvature.  

Thus  (\ref{surf1}) shows that  a generic orthonormal frame in $\mathbb R^3$ 
is transported along the surface by the following  equation
\begin{equation}
d  \left( \begin{matrix} { \bf x } \\ {\bf e}_1 \\ {\bf e }_2 \\ {\bf n} \end{matrix}
\right)  =  
\left( \begin{matrix} 0 & \omega_1 & \omega_2 & 0 \\
0 & 0  & {\omega_1}^2  &{\omega_1}^3  \\ 
0 & {\omega_2}^1  & 0  & {\omega_2}^3  \\
0 & {\omega_3}^1  &  {\omega_3}^2    & 0 \end{matrix} \right)  
\left( \begin{matrix}{\bf x} \\  {\bf e}_1  \\ {\bf e }_2 \\ {\bf n} \end{matrix} \right),
\label{chern1}
\end{equation}
where $d$ is the exterior derivative and the $\omega$'s are one-forms.

Then the integrability of (\ref{chern1}) implies that acting on (\ref{chern1}) the exterior derivative must remain nilpotent, that is,
\begin{equation*}
d^2 = 0
\end{equation*}
 leading to the following structure equations 
\begin{eqnarray}
d{\omega_1}^2 - {\omega_1}^3 \wedge {\omega_3}^2 \  & = &  \   0 \ \ = \  \ d{\omega_2}^1 -{ \omega_2}
^3\wedge{\omega_3}^1 ~~~~~~ 
\label{se3a}
\\
d{\omega_1}^3 - {\omega_1}^2 \wedge {\omega_2}^3 \  & = &  \ 0 \ \ = \ \ d{\omega_{3}}^1 - {\omega_3}^2 \wedge 
{\omega_2}^1 ~~~~~~  
\label{se3b} 
\\
d{\omega_2}^3 - {\omega_2}^1 \wedge {\omega_1}^3 \  & = &  \  0 \ \  = \  \ d{\omega_{3}}^2 - {\omega_3}^1 \wedge
{\omega_1}^2. ~~~~~~ 
\label {se3c}
\end{eqnarray}
In fact, equation (\ref{se3a}) is the Gau\ss~equation; while equation (\ref{se3b}) and (\ref{se3c}) are the Codazzi equations.  

In addition the following two equations (also) occur
\begin{eqnarray}
d\omega_1 - \omega_2  \wedge {\omega_2}^1  \  & = &  \   0 \  \ = \ \ d\omega_2 - \omega_1 \wedge  {\omega_1}^2 ~~~~
\label{se4a}
\\
\omega_1\wedge {\omega_1}^3 + \omega_2 \wedge {\omega_2}^3 \ & = & \ 0 %
\label{se4b}.
\end{eqnarray}

Then  equations (\ref{se3a})-(\ref{se4b}) determine the surface $\mathcal S$
since: 
A solution to the Gau\ss-Codazzi equations  specifies the 
frames ($\mathbf e_1, \mathbf e_2, \mathbf n$) at each point on the surface; 
while the one-forms $\omega_1$ and $\omega_2$ can be constructed
from (\ref{se4a}) and  (\ref{se4b}), in  terms of a solution to the Gau\ss-Godazzi equations. That way the surface $\mathbf x(u,v)$ is   obtained, up to translations and rotations in $\mathbb R^3$, by integrating  the equations (\ref{e1e21}) and  (\ref{e1e22}), that is,
\begin{eqnarray*}
\frac{\partial \mathbf x}{\partial u} \ = \ \omega_{1 u} \mathbf e_1 + \omega_{2 u} \mathbf e_2 \\ 
\frac{\partial \mathbf x}{\partial v} \ = \ \omega_{1 v} \mathbf e_1 + \omega_{2 v} \mathbf e_2.
\end{eqnarray*}

The effect of a local frame rotation in ($\mathbf e_1, \mathbf e_2$) around the normal vector $\mathbf n$ by an angle $\chi$  corresponds to a rotation of the one-forms:
\begin{equation}
\left( \begin{matrix} \omega_1 \\ \omega_2 \end{matrix} \right) \ \to \ 
\left( \begin{matrix} \cos\chi  & -\sin\chi \\ \sin\chi & \cos\chi \end{matrix} \right)
\left( \begin{matrix} \omega_1 \\ \omega_2 \end{matrix} \right)
\label{rotsurf}
\end{equation}
However, from  (\ref{se4a}) one obtains
\[
{\omega_2}^1 \ = \ - {\omega_1}^2 \ \to \ {\omega_2}^1 + d\chi 
\]
{\it i.e.}  ${\omega_2}^1 $ transforms like the  U(1) connection $C$ in (\ref{gaugeA}), with $\chi = 2\theta$,  under the frame rotations.
Also, (\ref{se3b}) and  (\ref{se3c}) imply that 
\[
\left( \begin{matrix} {\omega_3}^1  \\ {\omega_3}^2 \end{matrix} \right) \ \to \ 
\left( \begin{matrix} \cos\chi  & \sin\chi \\ -\sin\chi & \cos\chi \end{matrix} \right)
\left( \begin{matrix} {\omega_3}^1 \\  {\omega_3}^2 \end{matrix} \right)
\]
thus
\[
\Omega_\pm \ = \ {\omega_3}^1 \pm i {\omega_3}^2 \ \buildrel{\chi}\over{\longrightarrow} 
e^{\pm i\chi} \Omega_\pm
\]
transforms like the  complex one-form $B_\pm$ in (\ref{gaugeA}) 
under the U(1) rotation.

To sum up note  that (${\omega_2}^1, \Omega_\pm$) $\sim$ ($C,B_\pm$) can be interpreted as a two dimensional SU(2) multiplet, 
under the SO(2)$\sim$U(1) frame rotation (\ref{rotsurf}) around the normal vector
$\mathbf n$. 

Next note that the relation (\ref{se4b}) remains  invariant under this U(1)  rotation. Moreover, by choosing ($u,v$)
to be the coordinates in the principal directions at the point $\mathbf x(u,v)$
 one gets
\[
{\omega_3}^1 = \kappa_1 \omega_1, \ \ \ \ \ \ \ \ \ {\omega_3}^2 = \kappa_2 \omega_2
\]  
where $\kappa_1$ and $\kappa_2$ are the two principal curvatures of the surface; this is akin to choosing the Frenet frames in the case of strings. 
Then, the relation (\ref{se3a})  becomes
\begin{equation*}
dF \sim d{\omega_2}^1 \ = \ K {\omega_3}^1 \wedge {\omega_3}^2 
\end{equation*}
{\it i.e.} equation (\ref{RK}), implying that we have fully recovered the formalism of decomposed  SU(2) Yang-Mills theory/Abelian Higgs Model  in the case of generic two-dimensional  Riemann surfaces (for details see  section VIII.A.). 

Finally, observe that 
\[
\kappa_1 + i \kappa_2 \simeq \rho_1 + i\rho_2.
\]
This completes the identification of the structure of decomposed two dimensional SU(2) Yan-Mills theory, and
the Gau\ss-Godazzi construction of surfaces.

\subsection{Spinors and surfaces}

Next we proceed to transcribe (\ref{chern1}) into a spinorial form,  following Section III. By introducing the spinor (\ref{psi1}), the conjugate spinor  (\ref{C}) and following (\ref{tpsi}),  the spinor is related to the normal vector $\mathbf n$ since
\begin{equation*}
\mathbf n \ = \ <\psi, \hat{\sigma} \psi> \ = \ - <\bar \psi, \hat{\sigma} \bar \psi>.
\end{equation*}
  Note that the spinor $\psi$ is akin to the Gau\ss~map, it defines
a mapping from the surface $\mathcal S$ to a complex two-sphere.
Then for the tangent vectors ($\mathbf e_1, \mathbf e_2$) of the surface $\mathcal S$ following (\ref{e+}) and (\ref{e-}) one gets
\begin{equation*}
\mathbf e_1 = \frac{1}{2} <\bar \psi, \hat \sigma \psi>, \ \ \ \ \ \ \ \ \ \ \mathbf e_2 = \frac{1}{2} <\psi, \hat \sigma \bar\psi>.
\end{equation*}
The spinor version of (\ref{chern1}), in terms of (\ref{majspi}) (by suppressing the equation for $d\mathbf x$), becomes
\begin{equation*}
d\Psi  \ = \ d \left( \begin{matrix} -\bar \psi  \\ \psi \end{matrix} \right) \ = \ \left( \begin{matrix} \alpha & \beta \\ \gamma & \delta
\end{matrix} \right) \left( \begin{matrix} -\bar\psi \\ \psi \end{matrix} \right)
\end{equation*}
where the one-form valued matrix is defined as
\begin{equation*}
\left( \begin{matrix} \alpha & \beta \\ \gamma & \delta
\end{matrix} \right) \ = \ \left( \begin{matrix} <\bar\psi, d\bar\psi>  & - <\psi, d\bar\psi>  \\ - <\bar\psi, d\psi> & 
<\psi, d\psi>
\end{matrix} \right) .
\end{equation*}
This yields to the relations
\begin{eqnarray}
{\omega_2}^1   &= & - 2  \, {\rm Im}  < \psi, d\psi> 
\label{omegalpha}
\\
{\omega_3}^1 & = &   2 \, {\rm Re} <\bar \psi, d\psi>
\label {spinomega31}
\\
{\omega_3}^2 & = &   2 \, {\rm Im} <\bar \psi, d\psi> 
\label{spinomega32}
\end{eqnarray}
{\it i.e.} the  spinor version of the Gau\ss-Godazzi equations, derived by direct substitution of (\ref{omegalpha})-(\ref{spinomega32}) in (\ref{se3a})-(\ref{se3c}). 

Let us conclude by pointing out that in terms of the components (\ref{psi1}) one obtains 
\[
{\omega_2}^1 \ = \ i \left[ z_1^\star d z_1 - z_1 d z_1^\star + z_2^\star d z_2 - z_2 d z_2^\star \right]
\]
{\it i.e.} the standard  composite U(1) gauge field in the $\mathbb C \mathbb P^1$ model. 
Recall that, in terms of (\ref{para})
\[
{\omega_2}^1  \ = \ \cos\vartheta d\phi_- - d\phi_+
\]
one can  recognize the structure of the Dirac monopole connection.  Similarly, for $\Omega_\pm$  one gets
\begin{eqnarray*}
{\omega_3}^1 + i {\omega_3}^2   &= & 2 ( z_1 d {z_2} - z_2 d z_1 )\\
& = & \ \frac{1}{2} \left( \mp i d \vartheta - \sin \vartheta d \phi_- \right).
\end{eqnarray*}

\subsection{NLSE as a surface in $\mathbb R^3$}

In order to construct a Lax pair  that yields an integrable system,  a  deformation of the  Maurer-Cartan form (\ref{omegalpha})-(\ref{spinomega32}) needs to be introduced. In what follows, as an example
it is described how the NLSE equation and the Liouville equation arises as  a decomposed deformation.  

In \cite{sym}-\cite{sym2} several examples of integrable systems that describe surfaces in $\mathbb R^3$ have been presented.  
In the case of NLSE, let us start with the induced metric
\[
g_{\alpha \beta} \, dw^\alpha dw^\beta  = \left(du \right)^2 + \kappa^2 \left(dv \right)^2 \ = \ 
\partial_\alpha \mathbf x \cdot \partial_\beta \mathbf 
x \, d w^\alpha dw^\beta
\]
and choose  the second fundamental form to be
\[
h_{\alpha\beta}  = \left( \begin{matrix} \kappa  & - \ \lambda \kappa  \\ - \lambda \kappa & \frac{\kappa^3}{2}  \! -\! 
\kappa\eta \end{matrix} \right).
\]
Here ($\kappa, \lambda ,\eta$) are three functions that  decompose the second fundamental form, and
need to be specified. Following (\ref{e1e21}) and  (\ref{e1e22}) and  using the induced metric one can  identify as
\begin{eqnarray*}
\mathbf e_1 & = & \frac{\mathbf x_u}{|| \mathbf x_u ||}  =   \mathbf x_u \\
\mathbf e_2   & = &  \frac{\mathbf x_v}{|| \mathbf x_v ||}   =  \frac{1}{\kappa} \mathbf x_v.
\end{eqnarray*}
Then  comparing  (\ref{surf1}) and  (\ref{chern1}) one computes that
\begin{eqnarray*}
{\omega_1}^2 & = &   \kappa_u  dv \\
{\omega_3}^1 & = &  \lambda \kappa dv - \kappa du  \\
{\omega_3}^2 & = &  \lambda du - \left( \frac{1}{2} \kappa^2 - \eta \right) dv.
\end{eqnarray*}
By substituting  this in the Gau\ss~ and Codazzi equations (\ref{se3a})-(\ref{se3c}) one obtains
\begin{eqnarray}
\kappa_v & = & - \lambda_u \kappa - 2\lambda \kappa_{u} \label{rios1} \\
\eta & = & \frac{1}{\kappa} \left[ \kappa_{uu} - \lambda^2 \kappa + \frac{\kappa^3}{2} \right],
\label{rios2}
\end{eqnarray}
where, when  identifying 
\begin{eqnarray*}
\lambda & = & \partial_u \Phi \\
\eta & = & \partial_v \Phi  
\end{eqnarray*}
and 
\[
\Phi(u,v)  \ = \  \int^u du^\prime \, \tau(u^\prime, v) 
\]
the NLSE equation (\ref{darios}) is recovered by taking 
the derivative of ({\ref{rios2}) with respect to $v$.

\subsection{Isothermic surfaces and integrable models}

A correspondence between  two-dimensional isothermic manifolds and integrable models has been pointed out in \cite{sym}-\cite{burs2}.
In particular it has been shown that isothermic surfaces have a one-parameter family {\it i.e.} a {\it pencil} of flat connections. Thus  there exists a putative one-parameter family of Lax pairs, with the parameter corresponding to  the spectral parameter of integrable hierarchy. 

Let us  introduce isothermal coordinates so that the induced metric admits  the conformal form
\begin{equation*}
g_{\alpha\beta} = e^{\phi} (du^2 + dv^2) \ = \ \partial_\alpha \mathbf x \cdot \partial_\beta \mathbf 
x \, d u^\alpha du^\beta
\end{equation*}
and  decompose the second fundamental form as 
\begin{equation*}
h_{\alpha\beta} \ = \ \left( \begin{matrix}  H e^{2\phi} + \frac{1}{2}(Q + \bar Q) & \frac{i}{2} (Q - \bar Q) \\
\frac{i}{2} (Q - \bar Q) & H e^{2\phi} - \frac{1}{2}(Q + \bar Q) \end{matrix} \right).
\end{equation*}
Accordingly,
\[
g^{\alpha\beta} h_{\alpha\beta} \ = \ \kappa_1 + \kappa_2 \ = \ 2H
\]
{\it i.e.}  $H$ is the mean curvature;  and $Q$ is the Hopf differential. 
For an isothermal manifold the Hopf differential is real valued, so that akin to the metric tensor, the second fundamental form becomes diagonal
\begin{equation}
Q - \bar Q = 0. 
\label{QQbar1}
\end{equation}
Again, by   following  (\ref{e1e21}) and (\ref{e1e22}) and using the induced metric one can  identify 
\begin{eqnarray*}
\mathbf e_1 & = & \frac{\mathbf x_u}{|| \mathbf x_u||}  =   e^{-\phi} \mathbf x_u \\
\mathbf e_2   & = &  \frac{\mathbf x_v}{|| \mathbf x_v||}   =  e^{-\phi} \mathbf x_v.
\end{eqnarray*}
Then by comparing  (\ref{surf1}) and (\ref{chern1}) one computes 
\begin{eqnarray*}
{\omega_1}^2 & = &   - \frac{1}{2} \phi_v du + \frac{1}{2} \phi_u dv  \\
{\omega_3}^1 & = &  - \left[ \, e^{-\phi} H + e^{\phi}  Q\right]  \\
{\omega_3}^2 & = & - \left[ \, e^{-\phi} H - e^{\phi} Q\right]  dv.
\end{eqnarray*}
Finally by substituting  this in the Gau\ss~ and Codazzi equations (\ref{se3a})-(\ref{se3c}) one  obtains
\begin{equation}
\phi_{uu} + \phi_{vv} + e^{-2\phi} H^2 - e^{2\phi} Q^2 \ = \ 0.
\label{riosrios}
\end{equation}
Moreover,  in the case of a constant mean curvature surface for which the Hopf differential is also constant (and $H = Q$), 
the sinh-Gordon equation is derived while  if $Q=0$ the Liouville equation is obtained.


%
%
%
%
%
%
%
%
%
%
%
%


\section{Cartan Geometry, Surfaces in $\mathbb R^2$  and  Two Dimensional  Yang-Mills Theory}

In this section the Cartan geometry \cite{cart1}-\cite{wise} is used since it provides a unified framework to describe string, surfaces, integrable models and decomposed Yang-Mills theories.

\subsection{Cartan geometry }

In the Cartan approach in order to  describe a k-dimensional Riemannian manifold 
$\mathcal X$ one utilizes  a {\it model space},  {\it i.e.} a tangent space manifold $\mathcal M$  that can be more elaborate than the Euclidean $\mathbb R^k$  used in the Riemannian approach. 

Let us assume that  the model space $\mathcal M$ is a $k$-dimensional homogeneous coset 
\[
{\mathcal M}  \simeq  G/H.
\] 
In particular, consider
\[
{\mathcal M}  \simeq  SU(2)/U(1) \sim \mathbb S^2.
\]

Next consider the following  general approach:
The total space $G$ is a Lie group that acts transitively on $\mathcal M$ while the gauge group $H$  is the stabilizer subgroup.
 Accordingly the Lie algebra $\mathfrak g$ of $G$ is  resolved into the vector space sum, that is, 
\begin{equation*}
\mathfrak g \ = \ \mathfrak h \oplus \mathfrak m.
\end{equation*}
Here $\mathfrak h$ is the Lie algebra of $H$, and $\mathfrak m$ spans the linear tangent space of $\mathcal M$.
Then the Lie algebra $\mathfrak g$ has the following structure,
\begin{eqnarray*}
&&[ \mathfrak h , \mathfrak h ]   \ \subseteq   \ \mathfrak h  \\
&&[ \mathfrak h , \mathfrak m ]   \ \subseteq  \ \mathfrak m \\
&& [ \mathfrak m , \mathfrak m ]   \ \subseteq \  \mathfrak h + \mathfrak m.
\end{eqnarray*}

In addition, consider a Yang-Mills connection one-form $A$ with gauge group $G$,  defined on the Riemannian manifold $\mathcal X$. 
Then  the Lie algebra  $\mathfrak g$ valued connection can be decomposed into a linear combination as
\begin{equation}
A \ = \ \omega + e 
\label{caA}
\end{equation}
where $\omega$ is the spin connection of $\mathcal X$,  {\it i.e.} the projection of $A$ onto $\mathfrak h$. 
The co-frame $e$ of $\mathcal X$ is the projection of $A$ onto $\mathfrak m$.  

By  substituting the decomposition (\ref{caA}) in the curvature two-form one gets
\begin{equation}
F \ = \ dA+ A\wedge A \ = \ F_{\mathfrak h} +F_{\mathfrak m}
\label{caF}
\end{equation}
where $F_{\mathfrak h}$  is  the projection of $F$ onto $\mathfrak h$
\begin{equation}
F_{\mathfrak h}  = \mathcal R + (e\wedge e)_{| \mathfrak h}  = d\omega + 
\omega \wedge \omega + (e\wedge e)_{| \mathfrak h}
\label{Fh}
\end{equation}
while $F_{\mathfrak m}$  the projection onto $\mathfrak m$
\begin{equation}
F_{\mathfrak m}  = \mathcal T + \left(e\wedge e\right)_{| \mathfrak m}  =  de + \omega \wedge \omega +
(e\wedge e)_{| \mathfrak m}.
\label{Fs}
\end{equation}
Note that   (\ref{Fh}) and (\ref{Fs}) correspond to the Cartan structure relations when we identify  
$\mathcal R$ as the Riemann-Cartan curvature two-form and $\mathcal T$ as the torsion two-form of the manifold $\mathcal X$.

In addition,  (\ref{Fh}) and (\ref{Fs}) are covariant under gauge transformations in the 
subgroup $H$. In particular,   the  gauge transformation $h \in H$ acts as follows
\begin{eqnarray}
\omega & \to & h^{-1} \omega h + h^{-1} d h 
\nonumber
\\ 
 e &\to & h^{-1} e h 
\nonumber
\\
 F &\to & h^{-1} F h.
 \label{carh3}
\end{eqnarray}

Then the  Yang-Mills action on $\mathcal X$ becomes decomposed into
\begin{eqnarray}
\hspace{-15mm} S & = & \frac{1}{4} \int \mbox{tr} \left(F \wedge \star F\right)  \nonumber\\
& = & \frac{1}{4} \int \mbox{tr}  \left( F_{\mathfrak h} \wedge \star F_{\mathfrak h} +
 F_{\mathfrak m} \wedge \star F_{\mathfrak m} \right) \nonumber\\
& = & \frac{1}{4} \int \mbox{tr}  \! \left\{ 2 \, \mathcal R \wedge \star \left(e \wedge e\right)_{| \mathfrak h} + \left(e \wedge e\right)\wedge \star
\left(e \wedge e\right)\right. \nonumber\\
& & \ \ \ \left. + \  \mathcal R \wedge \star \mathcal R + \mathcal T \wedge \star \mathcal T + 2 \mathcal T \wedge \star 
(e \wedge e)_{| \mathfrak m} \right\}. \ \ \
\label{YMdecos}
\end{eqnarray}
Note that we tacitly assume that the Killing metric is non-degenerate, with $\mathfrak h$ and $\mathfrak m$
mutually orthogonal. 
We also note that the first term of (\ref{YMdecos}) has the functional form of the Einstein-Hilbert action in the Palatini
formalism, while the second term is akin to the cosmological constant contribution.

Explicitly, if  $f_{\alpha \beta}^\gamma$ for $\alpha, \beta, ... = 1,...,\dim(\mathfrak g)$ are the Lie algebra structure constants of $\mathfrak g$ and $r,s,... = 1,...,\dim(\mathfrak m)$ label the subspace $\mathfrak m$ and $a,b,... =
\dim(\mathfrak m) + 1,...,\dim(\mathfrak g)$ label the remaining $\mathfrak h$, the relations (\ref{Fh}) and  (\ref{Fs}) become
\begin{eqnarray*}
F^a_{\mathfrak h} &= & d\omega^a + f_{bc}^a \, \omega^b \wedge \omega^c + f_{rs}^a e^r \wedge e^s\\
F^s_{\mathfrak m} & = &  de^s  + f_{ar}^s \, \omega^a \wedge e^r + f_{rt}^s \, e^r \wedge e^t.
\end{eqnarray*}

Moreover, when  $A$ being the $\mathfrak g$ valued Maurer -Cartan form
\[
A = g^{-1} dg
\]
the total curvature vanishes, and the local geometry of $\mathcal X$ coincides with that of the model space $\mathcal M$. 
In this case  from (\ref{Fh})  the curvature two-form $\mathcal R$ of $\mathcal X$ can be obtained since 
\begin{equation*}
F_{\mathfrak h} \ = \ 0 \ \Rightarrow \ \mathcal R \equiv d\omega +  \omega \wedge \omega  = - (e\wedge e)_{| \mathfrak h} .
\end{equation*}
This implies in particular that $\mathcal X$  is locally conformally flat. 
Similarly, from (\ref{Fs}) the components of the torsion two-form $\mathcal T$ can be derived since
\begin{equation*}
F_{\mathfrak m} \ = \ 0 \ \Rightarrow \ \mathcal T \equiv de +  \omega \wedge e  = - (e\wedge e)_{| \mathfrak m} .
\end{equation*}

Next cosnider the specific case: $G$ = $SU(2)$. 
The model space is $\mathcal M = \mathbb S^2$, while  the Hopf fibration
\[
SU(2)/U(1) \simeq \mathbb S^3 /\mathbb S^1 \sim \mathbb S^2
\] 
coincides with the Dirac monopole bundle in $\mathbb R^3$. As before, we choose the SU(2) Lie algebra generators 
to be the standard Pauli matrices, and introduce the complex combinations $\sigma^\pm$ defined in   (\ref{gmat2}).  In this representation, 
\[
\mathfrak s \mathfrak u (2) \simeq \mathfrak u(1) \oplus T\mathbb S^2
\]
with $T\mathbb S^2$ the tangent bundle of $\mathbb S^2$, the Hopf fibration structure of the SU(2) Lie algebra is manifest since 
\begin{eqnarray*}
&& [ \mathfrak u(1) , T\mathbb S^2 ] \subseteq T\mathbb S^2 \\
&& [T\mathbb S^2 , T\mathbb S^2 ] \subseteq \mathfrak u(1)\\
&& [\mathfrak u(1), \mathfrak u(1)] \subseteq \mathfrak u(1)
\end{eqnarray*}
where $\sigma^3$ in $\mathfrak  u (1)$  and $\sigma^\pm$ the basis for $T\mathbb S^2 $.
Then the $SU(2)$ valued connection one-form $A$ is  on the manifold $\mathcal X$ (which we shall specify in more detail in the sequel). 
In line with (\ref{caA}) we decompose $A$ as 
\begin{equation*}
A \ = \ \omega \sigma^3 + e \sigma^+ + \bar e \sigma^- \ = \ \left( 
\begin{matrix} \omega & 0 \\ 0 & -\omega \end{matrix} \right) +
\left( \begin{matrix} 0  & e \\ -\bar e  &  0 \end{matrix} \right)
\end{equation*}
where  the off-diagonal part $e$ of $A$ in a complex basis are represented by the holomorphic polarization of the SU(2) Lie algebra.
Then, the Cartan curvature two-form $F$  decomposites as
\begin{eqnarray}
F& =& dA + A\wedge A \nonumber\\
&= &  \left( \begin{matrix}
d\omega - e\wedge \bar e & 0 \\ 0 &  - d\omega + e\wedge \bar e \end{matrix} \right)\nonumber\\
& + &\left( \begin{matrix}
0 & de + 2\omega \wedge e \\ -d \bar e - 2\omega \wedge \bar e  & 0  \end{matrix} \right).
\label{carF}
\end{eqnarray}

%
%
%
%
%
%
%

\subsection{Two dimensional Yang-Mills}

Finally, let us relate the previous construction to that of decomposed Yang-Mills, 
presented in section VIII.
Assume 
that  the manifold $\mathcal X$ is a Riemann surface in $\mathbb R^3$,   implying that (\ref{carF}) vanishes.
Then the vanishing of the $\sigma^3 \sim \mathfrak u (1)$ component of $F$ implies that   the curvature is 
\begin{equation}
\mathcal R = d\omega = e \wedge \bar e.
\label{carRee}
\end{equation}
Similarly,  the  $\sigma^\pm \sim T\mathbb S^2$ component states that the torsion vanishes, {\it i.e.}
\begin{equation*}
\mathcal T = de + \omega \wedge e = 0.
\end{equation*}
In the Cartan geometry, these two equations are the Bogomolny equations of (\ref{bogo}).

In addition, the integral 
\[
\frac{1}{4\pi} \int \mathcal R =  \frac{1}{4\pi} \int  e \wedge \bar e
\]
over a surface in $\mathbb R^3$ yields the Gau\ss-Bonnet formula that computes 
its Euler characteristic; see equation (\ref{Chernchar}) and the subsequent analysis

For example in order to use explicit formulas, let us introduce $g \in SU(2)$ to be of the form
\[
g = \left( \begin{matrix} \alpha & \beta \\ - \bar\beta & \bar\alpha \end{matrix} \right)
\]
where, motivated by the  Gau\ss ~map  (\ref{nvecp}) and (\ref{gausmap}),  set
\[
\alpha = \cos \frac{\vartheta}{2} e^{ \frac{i}{2} (\psi +\phi) } \ \equiv \ \cos\frac{\vartheta}{2} e^{ i \psi_+}
\]
\[
\beta = \sin \frac{\vartheta}{2} e^{ \frac{i}{2} (\psi - \phi) } \ \equiv \ \sin\frac{\vartheta}{2} e^{ i \psi_-}.
\]
In this case ($\vartheta, \psi_\pm$) can be identified as the  angles that parametrize the two hemispheres of  $\mathbb S^2 \in \mathbb R^3$ of the Gau\ss~map of the surface $\mathcal X$,  possibly with $n$-fold covering of the ensuing sphere.  The integer $n$ counts  the number of $2\pi$ circulations in $\psi_\pm$. 

Then   the Maurer-Cartan form, see (\ref{calb}),  (\ref{cala}) and (\ref{B+-}),  becomes
\begin{eqnarray}
\omega & = & \frac{1}{2} \left( \cos \vartheta \, d\psi + d\phi \right) 
\label{omegaYM}
\\
e & = & \frac{1}{2} e^{-i\phi} \left( - i d\vartheta + \sin \vartheta \, d\psi \right).
\label{eYM}
\end{eqnarray}
Thus the standard  Dirac (Hopf)  monopole bundle in $\mathbb R^3$ has been recovered, with $\omega$  being the monopole connection, while the monopole number coincides with $n$. 
Moreover, the {\it r.h.s.} of (\ref{carRee}) becomes
\[
e \wedge \bar e \ = \ - \frac{i}{2} \sin\vartheta \, d \vartheta \wedge \psi
\]
{\it i.e.}  the standard functional form of volume two-form on the two-sphere 
$\mathbb S^2$ of the Gau\ss~map.

%
%
%
%
%
%
%
%
%
%
%
%
%

\section{D=4 Yang-Mills  and embedded two dimensional integrable models}

In this section,   the four dimensional SU(2) Yang-Mills theory and its 
embedded   two dimensional models is studied.
 In particular, the two dimensional embedding structures are  related to strings, their dynamics, and Riemann surfaces in $\mathbb R^3$.  
The technique used is a  Kaluza-Klein reduction in combination of field variable elimination akin to the Hopf differential (\ref{QQbar1}).

Let us   consider a four dimensional manifold that has the product structure 
$\mathcal M \times \mathcal N$ where $\mathcal M$ and $\mathcal N$ are both two dimensional. 
The manifold $\mathcal M$ is the base manifold. When needed,  it is equiped with the local  coordinates ($x_1,x_2$)$\sim$($u,v$) which correspond to the 
space-time coordinates of the field variables  in the putative two dimensional integrable field theory. 
 The  Kaluza-Klein reduction  takes place with respect to the two dimensional auxiliary manifold $\mathcal N$  fibered over $\mathcal M$ (or {\it vice versa}).   
 
Then the imposed requirement  on $\mathcal M$  and $\mathcal N$ is 
that the four dimensional  product manifold $\mathcal M \times \mathcal N$ is
locally conformally flat.
{\bf Remark:} This geometric structure will be imposed upon us in the sequel,
by the structure of the decomposed four dimensional Yang-Mills theory.

$\bullet$ Examples with  compact $\mathcal M$  include
Riemann surfaces with genus $g \geq 1$,  that are fibered by the two-dimensional 
torus $\mathcal N = \mathbb T^2 = \mathbb S^1 \times \mathbb S^1$. 

$\bullet$ Examples of $\mathcal M$ with 
hyperbolic geometry  include warped products of the  
Poincar\'e half-plane $\mathbb  H^2$ fibered with 
manifolds such as $\mathbb R^2$, $\mathbb S^2$, $\mathbb T^2$, etc.
 
$\bullet$ Note that neither the product manifold $\mathbb S^2 \times \mathbb S^2$ nor the
product manifold $\mathbb S^2 \times \mathbb T^2$ is locally conformally flat. 

%
%
%
%
%
%

\subsection{Yang-Mills Field}
\vskip 0.2cm

Let us start by using  the holomorphic basis (\ref{gmat2}) of the $\mathfrak s \mathfrak u(2)$ Lie algebra, {\it i.e.}  the following  expansion of the connection one-form 
$A$:
\begin{equation}
A \ = \ A_\mu^3 \sigma^3 dx^\mu + X_\mu^+ \sigma^- dx^\mu + X_\mu^- \sigma^+ dx^\mu
\label{pauli2}
\end{equation}
where
\[
X_\mu^\pm = A_\mu^1 \pm i A_\mu^2.
\]
Similarly, to the two dimensional  Cartan's formalism  select the $SU(2)$ gauge group to be the total space $G$;
acting  on the gauge field in the usual fashion
\[
A \ \to \ gA g^{-1} + 2i \, gd g^{-1}.
\]
The gauge group $H$ is  chosen to be  the diagonal Cartan subgroup $H \simeq U_C(1) \in SU(2)$. 
In particular, by setting 
\[
h = e^{\frac{i}{2} \omega \sigma^3} \ \ \in \ U_C(1)
\]
one finds that the component $A^3_\mu \sim A_\mu$  transforms as a $U_C(1)$ gauge field
\[
\delta_h A_\mu \ = \ \partial_\mu \omega
\]
while the off-diagonal $X_\mu^\pm$ gives
\[
\delta_h X_\mu^\pm \ = \ e^{\mp i \omega} X_\mu^\pm.
\]
In parallel with (\ref{carh3})  the Yang-Mills field strength tensor 
\[
F_{\mu\nu}^a \ = \ \partial_\mu A_\nu^a - \partial_\nu A_\mu^a + \epsilon^{abc} A_\mu^b A_\nu^c \ \ \ \ (a=1,2,3)
\]
decomposes as follows:

 The diagonal Cartan component gives
\begin{eqnarray}
F_{\mu\nu}^3  &= & F_{\mu\nu} + P_{\mu\nu} \sim F_{\mathfrak h}\nonumber\\
&=& \partial_\mu A_\nu - \partial_\nu A_\mu + \frac{i}{2} (
X_\mu^+ X_\nu^- - X_\nu^+ X_\mu^-) 
\label{FPh}
\end{eqnarray}
where
\begin{eqnarray}
F_{\mu\nu} &= & \partial_\mu A_\nu - \partial_\nu A_\mu \nonumber \\
P_{\mu\nu} & = &  \frac{i}{2} (
X_\mu^+ X_\nu^- - X_\nu^+ X_\mu^-).
\label{Pmunu}
\end{eqnarray}
Finally, the off-diagonal components becomes
\begin{eqnarray*}
 F_{\mu\nu}^\pm 
&=&  F_{\mu\nu}^1 \pm i F_{\mu\nu}^2  \sim F_{\mathfrak m} \nonumber\\
&=& (\partial_\mu \pm i A_\mu) X_\nu^\pm
- (\partial_\nu \pm i A_\nu) X_\mu^\pm\\
 &=& \ D^\pm_{A\mu} X^\pm_\nu -  D^\pm_{A\nu} X^\pm_\mu.
\end{eqnarray*}

%
%
%
%
%
%
%
%
%

\subsubsection{Grassmannian structure}

The  antisymmetric tensor $P_{\mu\nu}$ in (\ref{Pmunu}) obeys the condition 
\[
P_{12}P_{34} - P_{13}P_{24} + P_{23}P_{14} \ = \ 0.
\]
In the context of projective geometry, this is the relation that 
defines the Pl\"ucker coordinates of the Klein quadric. 
The Klein quadric describes the embedding of the real Grassmannian $G(4,2)$ in the five dimensional projective space 
$\mathbb R \mathbb P^5$ as a degree four hypersurface. 
This Grassmannian is the four dimensional
manifold of two dimensional planes that are embedded in $\mathbb R^4$. It  coincides with the  homogeneous space
\begin{equation*}
G(4,2) \ \sim \ \frac{ SO(4) }{SO(2) \times SO(2) } \ \simeq \ \mathbb S^2 \times \mathbb S^2.
\end{equation*}

Let us describe  a two-plane in $\mathbb R^4$ with an orthonormal {\it zweibein}
\begin{equation*}
e_\mu^\alpha e_\mu^\beta \ = \ \delta^{\alpha\beta} \ \ \ \ (\alpha,\beta=1,2)
\end{equation*}
or  in terms of the complex base,
\begin{equation}
e_\mu = \frac{1}{\sqrt{2}} ( e^1_\mu + i e^2_\mu).
\label{realime}
\end{equation}
Note that
\begin{eqnarray*}
e_\mu e_\mu \ = \ 0, \ \ \ \ \ \ \ \ 
e_\mu \bar e_\mu \ = \ 1.
\end{eqnarray*}
Since  $e_\mu$ spans a two dimensional plane in four dimensions,
 an additional complex zweibein can be introduced¬
\begin{equation*}
m_\mu = \frac{1}{\sqrt{2}} \left(  m_\mu^1 + i  m_\mu^2\right).
\end{equation*} 
That way an orthonormal basis that spans the entire $\mathbb R^4$ is obtained. That is, 
\begin{equation*}
\begin{array}{cccc}
<\! e \, , e \! >  & =  & < \!  m \, , m \! >
& = \  \  < \!  e ,  m \! > \  \  =\  0 
\\
<\!  \bar e ,  e \! > & = & < \!  \bar m , m \! > 
\, & \hskip -1.8cm  = \ 1. 
\end{array}
\end{equation*}

Next by  combining the $(e_\mu^\alpha ,  m_\mu^\alpha)$ into a $SO(4)$ valued $4\times 4$ {\it vierbein} matrix
\begin{equation*}
{{\mathbbmss e}^\kappa}_\mu \ = \ ( e^1_\mu , e^2_\mu , m^1_\mu , m^2_\mu)
\end{equation*}
the ensuing $SO(4) \simeq SU(2) \times SU(2)$ Cartan  equation 
\begin{equation*}
d {{\mathbbmss e}^\kappa} + {\omega^\kappa}_\tau {\mathbbmss e}^\tau = 0
\end{equation*}
gives  the components of the Levi-Civita connection one-form 
${{\omega}^\kappa}_\tau$ in terms of $e$ and $m$. That is, 
\begin{eqnarray*}
\hspace{-5mm} d  e &= & < \!  \bar e , d e \! > e + < \!  \bar m , d  e \!>  m + < \! m \, , d e \!> \bar m 
\nonumber\\ 
\hspace{-8mm} d m &= & < \! \bar m , d m \! > m + < \! \bar e , d  m \!>  e + < \!  e \, , d  m \!>
\bar e. 
\end{eqnarray*}

Next let us  identify as
\begin{equation}
C = i < \!  \bar e , d e \! > 
\label{Cintu1}
\end{equation}
 the  $U_I(1)$ connection. Then  by introducing a local frame rotation on the two-plane ($e_\mu, 
 \bar e_\mu$)  that sends
\begin{equation}
e_\mu \ \to \ e^{-i\lambda} e_\mu
\label{intu1a}
\end{equation}
leads to 
\begin{equation*}
C \ \to \ C + d\lambda.
\end{equation*}

Similarly, the dual  connection
\begin{equation}
Q = i  \! < \!  \bar m , d m \!> 
\label{2Lam}
\end{equation}
 transforms like a $U(1)$ gauge field 
\[
Q \to Q + d\chi
\]
under  a rotation of the $( m^1_\mu , m_\mu^2)$ frame that spans the (tangent plane of) $\mathcal N$, {\it i.e.}
\[
m \to e^{-i\chi}  m.
\]

The remaining components of ${{\omega}^\kappa}_\mu$
\begin{eqnarray*}
\Phi^{+}_\mu & = & < \!  m \, , \partial_\mu  e \!>
\nonumber \\
\Phi^-_\mu & = & < \!  m \, , \partial_\mu \bar e\!>
\end{eqnarray*}
together with their complex conjugates
transform homogeneously under the rotations
of the $( e_\mu^1, e_\mu^2)$ and $( m_\mu^1 ,  m_\mu^2)$
basis vectors since
\[
\Phi^\pm_\mu \ \longrightarrow \ e^{ -i(\chi \pm \lambda)} \,
\Phi^\pm_\mu.
\]

Finally, by combining  the ${{\omega}^\kappa}_\mu$
into the two $SU(2) $ Lie-algebra valued one-forms,
\begin{equation*}
\begin{array}{ccc}
\left(Q + C\right)
\sigma^3 & + & \frac{1}{2i} \Phi^+ \sigma^+ + 
\frac{1}{2i} \left(\Phi^+\right)^*
\sigma^- 
\\~\\
\left(Q - C\right)
\sigma^3 & + & \frac{1}{2i}  \Phi^- \sigma^+ + \frac{1}{2i}
\left(\Phi^-\right)^* \sigma^- 
\end{array}
\end{equation*}
a direct computation reveals that
\begin{eqnarray}
d\left(Q \pm C\right) & = & \frac{i}{2} \Phi^\pm \wedge (\Phi^\pm)^* 
\nonumber \\ 
d \Phi^\pm & = &  - i \left(Q \pm C\right) \wedge \Phi^\pm 
\label{mceqs}
\end{eqnarray}
{\it i.e.} the  $SO(4)\simeq SU(2)\times SU(2)$ Maurer-Cartan structure equations.

\subsubsection{Grassmannian electric-magnetic duality}

Let us proceed, by defining the tensor
\begin{equation*}
H_{\mu\nu} \ = \ \frac{i}{2} \left( e_\mu \bar e_\nu - e_\nu \bar e_\mu \right)
\end{equation*}
and introducing the electric ($E_i$) and magnetic $(B_i)$ components of $H_{\mu\nu}$,
\begin{eqnarray}
E_i &=& \frac{i}{2} ( e_0^{} e_i^\star - e_i^{} e_0^\star)
\nonumber \\
B_i & = &  \frac{i}{2} \epsilon_{ijk} e^\star_j e^{}_k.
\label{EB}
\end{eqnarray}
They are subject to the following properties
\begin{eqnarray*}
\vec E \cdot \vec B &=& 0
\nonumber \\
\vec E \cdot \vec E + \vec B \cdot \vec B & =& \frac{1}{4}.
\end{eqnarray*}
Next, by introducing the selfdual and anti-selfdual  combinations
\begin{equation}
\vec s_{\pm} = 2 ( \vec B \pm \vec E)
\label{s+-}
\end{equation}
 two independent unit vectors are obtained that  determine two-spheres $\mathbb S^2_\pm$ in $\mathbb R^4$.

Note that, using (\ref{s+-}) and inverting  (\ref{EB})   the zweibein $e_\mu$ can be expressed as
\begin{equation}
e_\mu = \frac{1}{2} e^{i\xi} \,
\left( \sqrt{ 1-\vec s_+ \cdot \vec s_- } \ , 
\ \frac{ \vec s_+ \times \vec s_- + i ( \vec s_- - \vec s_+) }{\sqrt{
1 - \vec s_+ \cdot \vec s_- }} \right). 
\label{es}
\end{equation}
Here the phase factor $\xi$ that is not visible in $H_{\mu\nu}$,
is a section of the internal $U_I(1)$ bundle.
The definition
\begin{equation}
\sin \eta = \frac{1}{\sqrt 2} \sqrt{ 1 + \vec s_+ \cdot \vec s_-}
\la{7eta}
\end{equation}
leads to
\begin{eqnarray*}
\vec {E} & = & \sin \eta \cdot \vec p \nonumber \\
\vec { B} & = & \cos \eta \cdot \vec r
\end{eqnarray*}
where $\vec p$ and $\vec r$ are two orthogonal unit vectors. 
Together with their exterion product 
\[
\vec q = \vec r \times \vec p
\]  
one obtains an orthonormal triplet.

Also, the zweibein $e_\mu$  take the form
\begin{eqnarray*}
 \frac{1}{\sqrt{2}}( e_\mu^1 + i  e_\mu^2) & =&
 \frac{e^{i\xi}}{\sqrt{2}} \left( \begin{array}{c}
\sin \eta  \\
- i \vec p - \cos \eta \cdot \vec q \ \end{array} \right)\nonumber\\
&\buildrel{def}\over{=} &  \frac{e^{i\xi}}{\sqrt{2}}( \hat e_\mu^1 + i  \hat e_\mu^2)
\nonumber 
\end{eqnarray*}
and the dual zweibein take the form
\begin{eqnarray*}
m_\mu & = & 
\frac{e^{i\delta}}{\sqrt{2}} \left( \begin{array}{c}
\cos \eta \\
- i \vec r + \sin \eta \cdot \vec q \ \end{array} \right) \nonumber \\
&\buildrel{def}\over{=} &  \frac{e^{i\xi}}{\sqrt{2}}( \hat m_\mu^1 + i  \hat m_\mu^2).
\end{eqnarray*}
Here $\xi$  and $\delta$ are the phase for the frame rotation on the corresponding
two plane.

Next,  define a transformation $\mathcal R$ 
\begin{eqnarray*}
\mathcal R e_\mu & = & m_\mu
\nonumber \\
 \mathcal R \bar e_\mu & = & \bar m_\mu
\nonumber
\\
\mathcal R^2 &=& \mathbb I.
\end{eqnarray*}
 which is a duality transformation between the two planes 
that are spanned by $e_\mu$ and $ m_\mu$ in $\mathbb R^4$, respectively.
Then the action of $\mathcal R$  for $H_{\mu\nu}$ coincides with the action of  the Hodge duality
\begin{equation*}
H_{\mu\nu} \  \buildrel {\mathcal R} \over \longrightarrow \   \star H_{\mu\nu}
\end{equation*}
where
\[
\star H_{\mu\nu} \ = \  \frac{i}{2} (m_\mu \bar m_\nu - m_\nu \bar m_\mu).
\]
On the electric and magnetic components of $H_{\mu\nu}$ 
the action of $\mathcal R$  coincides with  the electric-magnetic duality transformation
\begin{equation*}
\mathcal R \left( \begin{array}{c} \mathcal B \\ \mathcal E \end{array} \right) \
 \ = \
 \left( \begin{array} {cc} 0 & 1 \\
1 & 0 \end{array} \right)\left( \begin{array}{c} \mathcal B
\\ \mathcal E \end{array} \right).
\end{equation*}

The two unit vectors (\ref{s+-}) are eigenvectors of $\mathcal R$ since 
\[
\mathcal R \vec{\mathit s}_\pm = \pm \vec{\mathit s}_\pm.
\]
Thus  $m_\mu$ can be expressed   in terms of $\vec {\mathit s}_\pm$,  up to the overall phase, by simply acting with
$\mathcal R$  in (\ref{es}), {\it i.e.}
\begin{equation*}
m_\mu = \frac{1}{2} e^{i\delta} 
\left( \sqrt{ 1+\vec s_+ \cdot \vec s_- } \ ,
\  \frac{ \vec s_- \times \vec s_+ - i ( \vec s_- + \vec s_+) }{\sqrt{
1 + \vec s_+ \cdot \vec s_- }} \right).
\end{equation*}
Finally, equations (\ref{intu1a}) and (\ref{2Lam}) lead to 
\begin{eqnarray}
C_{\mu\nu}& = & \partial_\mu C_\nu - \partial_\nu C_\mu \nonumber\\
&= &  \vec{\mathit s}_+ \cdot
\left(\partial_\mu \vec{\mathit s}_+ \times \partial_\nu \vec{\mathit s}_+ \right) +  
\vec{\mathit s}_-\cdot
\left(\partial_\mu \vec{\mathit s}_- \times \partial_\nu \vec{\mathit s}_- \right)+ \Sigma_{\mu\nu}(\lambda)\nonumber\\
\label{Css}\\
Q_{\mu\nu} &=& \partial_\mu Q_\nu - \partial_\nu Q_\mu\nonumber\\
&=& \vec{\mathit s}_+ \cdot
\left(\partial_\mu \vec{\mathit s}_+ \times \partial_\nu \vec{\mathit s}_+  \right) -  
\vec{\mathit s}_-\cdot
\left(\partial_\mu \vec{\mathit s}_- \times \partial_\nu \vec{\mathit s}_- \right)+ \Sigma_{\mu\nu}(\delta)\nonumber\\
\label{Lss}
\end{eqnarray}
with  last terms in (\ref{Css}) and (\ref{Lss})   akin to Dirac string contributions, 
\begin{eqnarray*}
\Sigma_{\mu\nu}(\lambda) \ = \ \left[\partial_\mu , \partial_\nu\right] \lambda\nonumber\\
\Sigma_{\mu\nu}(\delta)  \ = \ \left[\partial_\mu , \partial_\nu\right] \delta.
\end{eqnarray*}
Note that the first two $\vec {\mathit s}_\pm$ dependent terms in (\ref{Css})
and ({\ref{Lss}) are related to each other by the duality $\mathcal R$.

In what follows, consider  the definition (\ref{s+-}). If  $\vec E=0$ then $\vec{\mathit s}_+ = \vec{\mathit s}_-  =\vec{\mathit s}$ while
\begin{eqnarray*}
C_{\mu\nu} &=& 2  \vec{\mathit s} \cdot
\left(\partial_\mu \vec{\mathit s} \times \partial_\nu \vec{\mathit s}  \right) + \ 
\Sigma_{\mu\nu}(\lambda)\\
Q_{\mu\nu} &=& \Sigma_{\mu\nu}(\delta)
\end{eqnarray*}
{\it i.e.} $C_\mu$ can be interpreted  in terms of a connection for magnetic monopoles, 
and, similarly,  $Q_\mu$ is  a connection for magnetic strings.  
On the other hand,  by setting  $\vec {\mathit B} = 0$ in ({\ref{s+-}) 
then $\vec{\mathit s}_+ = \vec{\mathit s}_- =\vec{\mathit s}$ while
\begin{eqnarray*}
C_{\mu\nu}& = & \Sigma_{\mu\nu}(\lambda)\\
Q_{\mu\nu} & = & 2 \vec{\mathit s} \cdot
\left(\partial_\mu \vec{\mathit s} \times \partial_\nu \vec{\mathit s}  \right)+ \ 
\Sigma_{\mu\nu}(\delta).
\end{eqnarray*}
Thus $Q_\mu$ is a connection for electric monopoles and $C_\mu$ is a connection for electric strings.

Let us conclude, by introducing the following explicit realization
\[
\vec{\mathit s}_\pm \ = \ \left( \begin{array}{c} \cos \phi_\pm \, \sin\theta_\pm \\ 
\sin\phi_\pm \, \sin\theta_\pm \\
\cos \theta_\pm \end{array} \right)
\]
implying that 
\begin{eqnarray*}
C_\mu & = & - \cos\theta_+ d\phi_+ - \cos\theta_- d\phi_-  + d\lambda
\\
Q_\mu &= & - \cos \theta_+ d\phi_+ + \cos\theta_- d\phi_- + d\delta.
\end{eqnarray*}
Once more the functional forms  of the Dirac monopoles are resolved, except that (now)   both electric and magnetic  monopoles occur; see (\ref{cala})   and (\ref{omegaYM}). In addition, from (\ref{mceqs}) one obtains
\[
\Phi^\pm = \mp i \, e^{\pm i \sigma} \left( d\theta_{\pm} \mp i \sin\theta_\pm d \phi_\pm\right)
\] 
where $\sigma$ is a phase,  coinciding with (\ref{cala}) and  (\ref{eYM}). 

%
%
%
%
%
%
%
%
%
%
%
%

\subsubsection{Grassmannian Decomposition of  Yang-Mills Field }

Initially, the fields $X^\pm_\mu$ in (\ref{FPh}) are decomposed in accordance 
with the Grassmannian structure
\begin{equation}
X_\mu^\pm = A_\mu^1 \pm i A_\mu^2 = \psi_1 e_\mu + \psi_2 \bar e_\mu.
\label{Xeebar}
\end{equation}
Since (\ref{Xeebar}) are components of the SU(2) connection, the Grassmannian structure
becomes locally defined: 

The $\psi_\alpha$ are two complex line bundles over $\mathbb R^4$ 
and ($ e_\mu, \bar e_\mu $) spans a bundle of two dimensional planes. These planes are akin the
tangent bundle of the manifold $\mathcal M$. The Hodge dual ($ m_\mu, \bar m_\mu $)
has a natural identification as the tangent bundle of the manifold $\mathcal N$. Finally, the vector bundles
(\ref{s+-}) can be viewed as the corresponding (locally defined) Gau\ss~maps.

 Then, by substituting (\ref{Xeebar}) into $P_{\mu\nu}$ one gets
\begin{eqnarray}
P_{\mu\nu}& =& \frac{i}{2} \left( \, |\psi_1|^2 - |\psi_2|^2 \, \right) \, \left( e_\mu \bar e_\nu - e_\nu \bar e_\mu\right)\nonumber\\
&=& \frac{i}{2} \, \rho^2 \, t_3 \, \left( e_\mu \bar e_\nu - e_\nu {\bar e}_\mu \right) \nonumber\\
& \equiv  & \rho^2 \, t_3 \, H_{\mu\nu}.
\label{Pt}
\end{eqnarray}
Also, by   introducing   the three component unit vector
\begin{equation}
\mathbf t \ = \ \frac{1}{\rho^2} \left( \psi_1^\star \ \psi_2^\star \right) \hat \sigma  \left( \begin{matrix} \psi_1 \\ \psi_2 \end{matrix} \right) \ = \ \left( \begin{matrix} \cos \phi \, \sin \vartheta \\ \sin\phi \, \sin \vartheta \\ \cos \vartheta
\end{matrix} \right)
\label{tpsi}
\end{equation}
where $(\psi_1,\psi_2)$ are  the local coordinates
\begin{eqnarray}
 &&\begin{matrix} \psi_1 & = &  \rho \, e^{i\xi} \cos \frac{\vartheta}{2} \, e^{-i\phi/2} \\
\psi_2 & = & \rho \, e^{i\xi} \sin \frac{\vartheta}{2} \, e^{i\phi/2},
\end{matrix} \label{psipara} \\
&&\rho^2 = |\psi_1|^2 + |\psi_2|^2
\label{confG}
\end{eqnarray}
 the tensor $H_{\mu\nu}$  satisfies the relation
\begin{equation*}
H_{\mu\nu}H_{\mu\nu} = \frac{1}{2}.
\end{equation*}

The decomposition (\ref{Xeebar}) entails an internal $U_I(1)$ symmetry, not visible to $A_\mu^a$, of the form
\begin{equation}
U_I(1) \ : \ \ \ \ \begin{matrix} e_\mu & \to & e^{-i\lambda} e_\mu \\
\psi_1 & \to & e^{i\lambda} \psi_1 \\
\psi_2 & \to & e^{-i\lambda} \psi_2. \\
\end{matrix}
\label{intu1}
\end{equation}
This  is the  local rotation (\ref{intu1a})   of the frames ($e_\mu^1 , e_\mu^2)$ that  coincides with 
the two dimensional plane of the Grassmannian. 
The ensuing connection is given by (\ref{Cintu1}) since
 \[
C_\mu \ \to \   C_\mu + \partial_\mu \lambda. 
\]
Finally, on the unit vector (\ref{tpsi}) the internal $U_I(1)$ transformation acts as 
\begin{equation}
t_{\pm} \ = \ \frac{1}{2} (t_1 \pm i t_2) \ \to \ e^{\mp 2 i \lambda} t_\pm
\label{tvecpm}
\end{equation}
while the component $t_3$ remains intact. Thus, $t_3$ is the projection of $\mathbf t$ onto the normal of the Grassmannian two-plane ($e_\mu^1, e_\mu^2$) towards the direction of a ``Gauss~map".  

Next let us define the $U_C(1) \times U_I(1)$ covariant derivative $\mathcal D_\mu$ as
\begin{eqnarray*}
\mathcal D_\mu \psi_1 & = & \left(\partial_\mu + i A_\mu - i C_\mu\right) \psi_1 \\
\mathcal D_\mu \psi_2 & = & \left(\partial_\mu + i A_\mu + i C_\mu\right) \psi_1 \\
\mathcal D_\mu e_\nu & = & \left(\partial_\mu + i C_\mu\right) e_\nu.
\end{eqnarray*}
On the components of the vector $\mathbf t$ it acts in the following manner:
\begin{equation*}
(\mathcal D_\mu)^{ab}  \  = \ \delta^{ab} \partial_\mu + 2 \epsilon^{ab3} C_\mu.
\end{equation*}

In addition of the continuous $U_I(1)$ transformation, the decomposition (\ref{Xeebar}) introduces the following discrete $\mathbb Z_2$ symmetry, that is,
\begin{equation*}
\mathbb Z_2 \ : \ \ \ \ \ 
\begin{matrix} 
e_\mu & \to & \bar e_\mu  \\
\psi_1 & \to & \psi_2 \\ 
\psi_2 & \to & \psi_1 \\  
C_\mu & \to & - C_\mu
\end{matrix}
\end{equation*}
while for the vector $\mathbf t$ one gets
\[
\mathbb Z_2 \ : \ \ \ \ \  \left( \begin{matrix} t_1 \\ t_2 \\ t_3 \end{matrix} \right) \ \to \ \left( \begin{matrix} t_1 \\ - t_2 \\ - t_3
\end{matrix} \right).
\]
This changes the orientation on the two-plane of the Grassmannian  spanned by $e_\mu$.
In terms of the angular variables (\ref{psipara}) it corresponds to
\[
\mathbb Z_2 \ : \ \ \ \ \  \begin{matrix} 
\vartheta &  \to & \pi-\vartheta \\
\phi &  \to &  2\pi -\phi.
\end{matrix}
\]
Thus, we may eliminate the $\mathbb Z_2$ degeneracy by a restriction to the upper hemisphere $\vartheta \in [0,\pi/2 ]$ of  $\mathbb S^2$. 

Note that under the $\mathbb Z_2$ transformation the Pl\"ucker coordinates $P_{\mu\nu}$
of (\ref{Pmunu}) remains intact. However, although under the continuous $U_I(1)$ rotation $H_{\mu\nu}$ is invariant, for $H_{\mu\nu}$ in (\ref{Pt}) one gets
\[
\mathbb Z_2 \ : \ \ \ \ \  H_{\mu\nu} \ \to \ - H_{\mu\nu}.
\]
At regular points where ($e^1_\mu, e^2_\mu$) determines a (co-)frame {\it i.e.} span  two planes,  the matrix $H_{\mu\nu}$ is non-degenerate. On the other hand, at points where $H_{\mu\nu}$ becomes
degenerate by setting either $\rho = 0$ or
$\vartheta = \pi/2$  the $P_{\mu\nu}$ in (\ref{Pt}) remains regular.  

Finally, the vector $\mathbf t$ by setting set $\vartheta = \pi/2$ simplifies to
\[
\mathbf t \ \to \ \left( \begin{matrix} \cos\phi \\ \sin\phi \\ 0 \end{matrix} \right).
\]
This corresponds to the boundary of the upper hemisphere that was introduced in order to eliminate the $\mathbb Z_2$ degeneracy.

%
%
%
%
%
%

\subsubsection{Grassmannian Decomposition of  Yang-Mills Lagrangian}

Recall that the Yang-Mills Lagrangian is defined as
\[
L_{YM} \ = \ \frac{1}{4} \left(F^a_{\mu\nu}\right)^2 + \frac{\zeta}{2} | D^+_{A\mu} X^+_\mu |^2.
\]
The second term is a gauge fixing term with  $\zeta$ being a gauge parameter. 

Let us, for simplicity,  choose the $\zeta \to \infty$ gauge. This is the widely used Maximal Abelian Gauge (MAG) 
\begin{equation*}
(\partial_\mu \pm i C_\mu) X^\pm_\mu  \ = \ D^\pm_{A\mu} X^\pm_\mu \ = \ 0.
\end{equation*}

When substituting the decomposition (\ref{pauli2}), (\ref{Xeebar}) and (\ref{tpsi}) in the Lagrangian one gets
\begin{equation}
L_{YM} \ = \ \frac{1}{4} ( F_{\mu\nu} + 2\rho^2 t_3 H_{\mu\nu})^2 
- \frac{3}{8} t_3^2 \rho^4 + \frac{1}{2} | D^+_{A\mu} X^+_\nu |^2.
\label{YMdec1}
\end{equation}
Comparison with  (\ref{caF})-(\ref{Fs}) leads to the following identifications terms,
\begin{eqnarray*}
\frac{1}{4} \mathcal R \wedge \star \mathcal R \ & \sim & \ \frac{1}{4} F^2_{\mu\nu}
\nonumber\\ 
\frac{1}{4} (e\wedge e) \wedge \star (e \wedge e) \ & \sim & \  \frac{1}{8} t_3^2 \rho^4 
\nonumber\\ 
\frac{1}{2} \mathcal R \wedge \star ( e \wedge e)_{| \mathfrak h}   \ & \sim & \ 
\rho^2 t_3 F_{\mu\nu} H_{\mu\nu} 
\end{eqnarray*}
and 
\begin{equation}
\frac{1}{4} \mathcal T \wedge \star \mathcal T \ \sim \ \frac{1}{2} | D^+_{A\mu} X^+_\nu |^2.
\label{YMact1}
\end{equation}
Note that the last term in (\ref{YMdecos}) is absent. 

%
%
%
%
%
%

\subsection{Torsion Term}
\vskip 0.2cm

Let us start by  analyzing the terms in (\ref{YMdec1}), one-by-one.
First, note that  the torsion term (\ref{YMact1}) is equal to 
\begin{eqnarray}
|\mathrm D^+_{A  \mu} X^+_\nu|^2 \!\!&=&\!\! 
|\mathcal D_{\mu} \psi_1|^2 + |\mathcal D_{ \mu} \psi_2|^2 +\rho^2 |\mathcal D_{\mu} e_\nu |^2\nonumber \\
\!\!&+&\!
  \frac{1}{2} \rho^2 t_+ 
\left(\bar{\mathcal D}_{\mu} \bar e_\nu\right)^2
+ \frac{1}{2} \rho^2 t_-  \left(\mathcal D_{\mu} e_\nu\right)^2.
\label{YM2nd}
\nonumber\\
\end{eqnarray}
Also, let us introduce the $U_C(1) \times U_I(1)$ invariant super-current 
\begin{eqnarray*}
2i |\psi_1 \psi_2 | J_\mu 
=  \left\{ \psi_1^\star \mathcal D_\mu \psi_1 - 
\psi_1 \bar{\mathcal D}_\mu \psi^\star_1\right.
\left. -\psi_2^\star \mathcal D_\mu \psi_2 -
\psi_2 \bar{\mathcal D}_\mu \psi_2^\star 
\right\} &&\nonumber \\
\Leftrightarrow \ |\psi_1 \psi_2 | J_\mu  \buildrel{def}\over{=} \ \rho^2(A_\mu - \partial_\mu \xi) - t_3 K_\mu. \hspace{28mm} &&
\end{eqnarray*}
Note that $\xi$ comes from the explicit  paramet\-rization (\ref{psipara}).
Then  the following 'tHooft tensor structure exists
\begin{eqnarray*}
T_{\mu\nu} & \buildrel{def}\over{=}& \partial_\mu (t_3 K_\nu) - \partial_\nu (t_3 K_\mu)\\
&=& \partial_\mu (t_3 C_\nu) - \partial_\nu (t_3 C_\mu) - \frac{1}{2} \mathbf t \cdot \partial_\mu \mathbf t \times 
\partial_\nu \mathbf t.
\end{eqnarray*}

Next by  introducing the linear combinations
\begin{eqnarray*}
\left( \begin{matrix} \mathcal J_\mu \\ \mathcal K_\mu \end{matrix} \right) &= & 
\left( \begin{matrix}  2\sqrt{t_+t_-} & t_3 \\ -t_3 & 2\sqrt{t_+t_-}  \end{matrix} \right)
\left( \begin{matrix} J_\mu \\  K_\mu \end{matrix} \right)\nonumber\\
&\equiv & 
\left( \begin{matrix}  \sin\vartheta & \cos\vartheta \\ - \cos\vartheta & \sin\vartheta  
\end{matrix} \right)
\left( \begin{matrix} J_\mu \\  K_\mu \end{matrix} \right)
\end{eqnarray*}
 one gets
\begin{eqnarray*}
 |\mathcal D_{\mu} \psi_1|^2 + |\mathcal D_{ \mu} \psi_2|^2
& =&  
\frac{1}{2} (\partial_\mu \rho)^2
+ \frac{\rho^2}{2}  (\partial_\mu \vartheta)^2 \nonumber \\
&&+ 
\frac{\rho^2}{2}  \sin^2 \vartheta \, (\mathcal J_\mu^2 
+ \mathcal K_\mu^2) ,
\end{eqnarray*}
while the following two relations are satisfied
\begin{eqnarray}
|\mathcal D_\mu \mathbf t|^2 &= &\left(\partial_\mu \vartheta\right)^2 + \left(1-t_3^2\right)  K_\mu^2\nonumber\\
&\equiv & 
\left(\partial_\mu \vartheta\right)^2 + \sin^2\vartheta  \, K_\mu^2
\nonumber\\
F_{\mu\nu}& \equiv& \partial_\mu A_\nu - \partial_\nu A_\mu \nonumber\\
& = & \partial_\mu \mathcal J_\nu - \partial_\nu \mathcal 
J_\mu  + \frac{1}{2} [\partial_\mu , \partial_\nu] \xi.
\label{Ftorsion}
\end{eqnarray}

Finally, let us consider  the last three terms in the torsion contribution (\ref{YM2nd}). Using the real and imaginary components of (\ref{realime}) one gets
\begin{eqnarray}
&&\rho^2 |\mathcal D_{\mu} e_\nu |^2 + \frac{1}{2} \rho^2 t_+ 
(\bar{\mathcal D}_{\mu} \bar e_\nu)^2
+ \frac{1}{2} \rho^2 t_-  (\mathcal D_{\mu} e_\nu)^2
=  \nonumber\\
&&\frac{\rho^2}{2} g_{\alpha \beta} ( \hat{\mathcal D}_\mu e_\nu)^\alpha (\hat{\mathcal D}_\mu e_\nu)^\beta, \ \ \ \ \ \ \ \ \ \   a,b=1,2.
\label{pseud1}
\end{eqnarray}
Using the frame rotation  (\ref{intu1}) and (\ref{tvecpm})] the covariant derivative and the metric tensor are defined by
\begin{eqnarray}
\hat{\mathcal D}_\mu^{ab} &= & \delta^{ab} \partial_\mu
+ \epsilon^{ab} (\cos \vartheta \, \mathcal J_\mu + \sin \vartheta
\, \mathcal K_\mu) \nonumber\\
g_{\alpha\beta} & = &
\left( \begin{matrix} 1+t_1 & t_2\\ t_2 & 1-t_1 \end{matrix} \right),
\label{metricsg}
\end{eqnarray}
while  the  Yang-Mills Lagrangian remains intact under this rotation. 

Finally, by choosing  
\[
2\lambda = \phi - \frac{\pi}{2}
\]
which sends $t_1 \to 0$ according to (\ref{tvecpm}) and (\ref{tpsi}), the metric (\ref{metricsg}) becomes
\[
g_{\alpha\beta} \ \to  \ 
\left( \begin{matrix} 1  & \sin\vartheta  \\ \sin\vartheta & 1 \end{matrix} \right)
\]
{\it i.e.} the metric  tensor of a pseudosphere with constant 
negative Gaussian curvature -2, in Chebyshev coordinates on the plane $\mathbb R^2$  
\begin{equation}
ds^2 = du_1^2 + du_2^2 + 2 \sin\vartheta \, du_1 du_2
\label{pseudos}
\end{equation}
when $\vartheta$ is a solution of the two-dimensional integrable 
sine-Gordon equation.

\subsection{Embedded Integrable Structures}

Initially, let us express the first term  of the Yang-Mills Lagrangian (\ref{YMdec1}) in terms  of the super-currents 
\begin{equation}
\frac{1}{4} \left( F_{\mu\nu} + 2\rho^2 t_3 H_{\mu\nu} \right)^2
=\ \frac{1}{4} \left( \partial_\mu \ \mathcal J_\nu - \partial_\nu 
\mathcal J_\mu+ 2\rho^2 t_3 H_{\mu\nu} \right)^2.
\label{firstYM}
\end{equation}
For simplicity, we overlook the Dirac string contribution in (\ref{Ftorsion}).
Also, let us point out the similarity between (\ref{firstYM}) and the first term in (\ref{Frho}) since
\begin{equation*}
\partial_\mu \ \mathcal J_\nu - \partial_\nu 
\mathcal J_\mu+ 2\rho^2 t_3 H_{\mu\nu} 
\sim  \partial_\alpha C_\beta - \partial_\beta C_\alpha - \left[1-\bar \rho \rho\right] H_{\alpha \beta}.
\end{equation*}

Then the  $H_{\mu\nu}$ contribution in  (\ref{firstYM}) gives rise to
\[
\frac{1}{4} P_{\mu\nu}^2 \ = \ \frac{1}{2} t_3^2 \rho^4
\]
which we combine  with the middle term of (\ref{YMdec1}). 
In addition, by  removing  all vector  fields in the Yang-Mills Lagrangian;  this is the analog of  
the  steps in (\ref{QQbar1})-(\ref{riosrios}), the following  structure is found to be 
embedded in the Yang-Mills Lagrangian
\[
\frac{\rho^2}{2} \left[ \left( \partial_\mu \vartheta \right)^2 + \frac{\rho^2}{4} \cos^2 \vartheta \right].
\]
Note that in  the London limit where $\rho$ is constant, 
and with a straightforward Kaluza-Klein reduction with $\mathcal N \simeq \mathbb R^2$  so that 
$\mathbb R^4 \mapsto \mathcal M \simeq \mathbb R^2$,   the integrable
sine-Gordon Lagrangian is obtained. Thus, we have identified the 
integrable structure of the sine-Gordon hierarchy together with the ensuing 
pseudosphere geometry (\ref{pseudos}).  Note that, both are naturally embedded in
the structure of the decomposed four dimensional Yang-Mills theory. 

Consider now the first term in the {\it r.h.s} of (\ref{pseud1}) which can be written as\begin{eqnarray*}
\rho^2 |\mathcal D_\mu e_\nu |^2 
&=&  \frac{1}{2} \left(\partial_\mu \eta\right)^2 
+ \frac{1}{2} (\partial_\mu \vec p)^2 \nonumber\\
&+&
\frac{1}{2} \cos^2 \! \eta \, \left\{  (\partial_\mu \vec q)^2  - 2 <\vec p , \partial_\mu \vec q>^2 \right\}.
\end{eqnarray*}
Then by defining
\begin{equation*}
\left( \begin{matrix} \vec u \\ \vec v \end{matrix} \right) \ = \ \left( \begin{matrix} \sin \eta & \cos\eta  \\ 
-\cos \eta & \sin \eta \end{matrix} \right)
\left( \begin{matrix} \vec p \\ \vec r \end{matrix} \right)
\end{equation*}
one gets
\begin{equation}
\rho^2 |\mathcal D_\mu e_\nu |^2 = \frac{1}{4} \rho^2 | \partial_\mu ( \vec u + i \vec v) |^2
=   \frac{\rho^2}{2} \left[ (\partial_\mu \vec E)^2 + (\partial_\mu \vec B)^2 \right].
\label{10fin}
\end{equation}
In this form, the invariance under the electric-magnetic duality transformation is manifest. 

By specifying  $\eta=0$ the purely magnetic contribution is obtained 
 \begin{equation*}
 \rho^2 |\mathcal D_\mu e_\nu |^2 \  \buildrel{\eta = 0}\over{\longrightarrow} \ \frac{\rho^2}{2} \left(\partial_\mu \vec r\right)^2
 \ = \ \frac{\rho^2}{2} (\partial_\mu \vec B)^2
\end{equation*}
which  leads to   the Heisenberg spin chain action in the vector $\vec r$.
Recall that the Heisenberg spin chain defines an integrable model in two dimensions,   which is also  a conserved charge in the NLSE hierarchy; see (\ref{ham1}).

Similarly, by specifying $\eta = \pi/2$ the purely electric contribution is obtained 
 \begin{equation*}
 \rho^2 |\mathcal D_\mu e_\nu |^2 \  \buildrel{\eta = 0}\over{\longrightarrow} \ \frac{\rho^2}{2} (\partial_\mu \vec p)^2
 \ = \ \frac{\rho^2}{2} (\partial_\mu \vec E)^2
\end{equation*}
which leads to the Heisenberg spin chain action in the vector field $\vec p$.
Therefore, the Heisenberg spin chain and the ensuing NLSE hierarchy, 
is embedded in  the decomposed representation of the four dimensional SU(2) Yang-Mills theory. Moreover,  the two variants are related to each other by electric-magnetic duality.

The remaining contribution to (\ref{pseud1}) can also be presented in terms of the vectors ($\vec E, \vec B$).
But these terms are also multiplied by the $U_I(1)$ dependent components $t_\pm$. Since the ground state  is necessarily $U_I(1)$ invariant and since $t_\pm$ transform according to (\ref{tvecpm}) the following conditions needs to be imposed
\[
t_\pm = 0.
\]

Following (\ref{bogo})  and  by assuming   the London limit ({\it i.e.} $\rho=0$ )  where the vector $\vec t$ acquires its ground state value leads to
\[
 \partial_\mu \mathcal J_\nu - \partial_\nu \mathcal 
J_\mu  + \frac{1}{2} [\partial_\mu , \partial_\nu] \xi + 2\rho^2 H_{\mu\nu} = 0
\]
 {\it i.e.} the  complexified Heisenberg model (\ref{10fin}). 

To conclude, note that with a Kaluza-Klein 
reduction of $\mathcal M \times \mathcal N \simeq \mathbb R^4 \mapsto \mathbb R^2$ and by specifying either to the electric ($\vec B = 0$) or to the magnetic ($\vec E=0$) sector equation (\ref{ham1}) is obtained; which corresponds to  an embedded integrable NLSE hierarchy within the decomposed 
four dimensional SU(2) Yang-Mills theory.

%
%
%
%
%
%
%
%
%
%
%
%

\subsection{Conformal Geometry}

Finally, we reveal  the local conformal geometry of the four dimensional Yang-Mills theory, alluded to in
the beginning of the present section \cite{ludvig}, \cite{chernolud}. 
Interpret (\ref{confG})  as the scale  of a (locally) conformally flat metric tensor
\begin{equation}
G^{}_{\mu\nu} \ = \ \left( \displaystyle{ \frac{\rho}{\Delta}} \right)^2
\delta^{}_{\mu\nu},
\label{met1} 
\end{equation}
where $\Delta$ is a constant with dimensions of mass and the parameter $\rho$ 
has dimensions of mass. Therefore, a dimensionful parameter is necessary
for the components of the metric tensor in order to acquire the correct dimensionality. 

Let us  introduce the vierbein as usual,
\begin{equation*}
G^{} _{\mu\nu}
\ = \ \delta_{ab}\,
{E^a}_\mu
{E^b}_\nu 
\end{equation*}
so that
\begin{eqnarray*}
{E^a}_\mu &= & \frac{\rho}{\Delta}\,
{\delta^a}_\mu \nonumber\\
{E^a}_\mu\, {E_b}^\mu& =& 
\delta^a_{\, \, \, \, b}.
\end{eqnarray*}

Then the  Christoffel symbol of the metric (\ref{met1}) takes the form
\begin{eqnarray*}
\Gamma^\mu_{\nu \sigma}& = & \frac{1}{2} G^{\mu\eta}(
\partial_\nu G^{}_{\eta \sigma} + \partial_\sigma G{}_{\eta \nu } -
\partial_\eta G^{}_{\nu\sigma  } )\nonumber\\
&=& \ \frac{1}{4} \{ \, \delta^\mu_\sigma \delta^\tau_\nu
+ \delta^\mu_\nu \delta^\tau_\sigma - \delta^{\mu\tau} 
\delta_{\nu\sigma} \, \} \partial_\tau \ln \sqrt{G}
\end{eqnarray*}
where
\[
\sqrt{G} =  \left( \displaystyle{ \frac{\rho}{\Delta}} \right)^4.
\]

Also, the spin connection $\omega^{\, \, a}_{\mu \, \, b}$ is obtained from
\[
\partial_\mu {E_a}^\nu + \Gamma_{\mu\lambda}^\nu {E_a}^\lambda -
\omega^{\, \, b}_{\mu \, \, a} {E_b}^\nu = 0
\]
which  gives
\begin{eqnarray*}
\omega^{\, \, a}_{\mu \, \, b} &= & {E^a}_\nu \nabla_\mu {E_b}^\nu \ = \
{E^a}_\nu \left( \partial_\mu {E_b}^\nu + \Gamma^\nu_{\mu\lambda} {E_b}^\lambda\right)
\nonumber\\
 &= & - {E_b}^\nu \nabla_\mu {E^a}_\nu = 
- {E_b}^\nu \left( \partial_\mu {E^a}_\nu - \Gamma^\lambda_{\mu\nu} {E^a}_\lambda\right).
\end{eqnarray*}
Note that these relations  define the action of the covariant  derivative $\nabla_\mu$ on the vector fields and the co-vector fields.
Explicitly, the spin connection is
\[
\omega^{\, \, a}_{\mu \, \, b} \ = \ \frac{1}{4}\left \{ \, 
{\delta^a}_\mu {\delta_b}^\sigma -
\delta_{bd} \, {\delta^d}_\mu \, \delta^{ac} {\delta_c}^\sigma \,
\right \} \partial_\sigma
\ln \sqrt{G}.
\]

Next let us employ the vierbein ${E^a}_\mu$ and the complex Grassmannian
zweibein (\ref{realime}) in order to introduce the following  complex zweibein
\begin{eqnarray*}
{\mathbbmss e}^{}_\mu &=& {E^a}_\mu  e^{}_a \\
{\bar\ce}_\mu& =& {E^a}_\mu \bar{e}_a.
\end{eqnarray*}
This zweibein is normalized {\it w.r.t.} the metric $G_{\mu\nu}$ according to
\begin{eqnarray*}
G^{\mu\nu} {\mathbbmss e}^{}_\mu \bar{\mathbbmss e}^*_\nu & = & 1 \\
G^{\mu\nu} {\mathbbmss e}^{}_\mu {\mathbbmss e}^{}_\nu 
= G^{\mu\nu} \bar{\mathbbmss e}^*_\mu \bar{\mathbbmss e}^*_\nu &=& 0.
\end{eqnarray*}
Initially, we push forward the spin connection into
\[
{\omega}^{\,\, a}_{\mu \, \, b} \ \to \
{\omega}^{\, \, \lambda}_{\mu \,\, \nu} \ 
= \  {E_a}^\lambda \, 
\omega^{\, \, a}_{\mu\,\, b} {E^b}_\nu
\]
then   introduce the covariantization of the internal $U_I(1)$ connection (\ref{Cintu1}) 
\begin{eqnarray}
{C}_\mu  & = &
i \bar\ce^\sigma ( \partial_\mu \ce^{}_\sigma - \Gamma^\lambda_{\mu\sigma}
\ce^{}_\lambda + \omega^{\, \, \lambda}_{\mu \,\, \sigma}\ce^{}_\lambda )\nonumber\\
&=& \ i {\bar \ce}^\sigma \nabla_\mu \ce_\sigma
+ i \bar\ce^{\lambda} \omega^{\,\, \sigma}_{\mu \, \, 
\lambda}  \ce^{}_\sigma
\label{covC}
\end{eqnarray}
and finally, we twist the covariant derivative operator with (\ref{covC}) as
\begin{equation*}
\nabla^{C}_\mu \ = \ \nabla_\mu + i C_\mu.
\end{equation*}

Finally,  the Yang-Mills Lagrangian can be recast  in a generally covariant format.  This results to
\begin{eqnarray}
L_{YM}\! \! &=&  \!\!\sqrt{G} \, \left\{ \frac{1}{4} G^{\mu\rho} G^{\nu\sigma} \mathcal F_{\mu\nu}
\mathcal F_{\rho\sigma} + \frac{\Delta^2}{12}
R -  \frac{3}{8} \Delta^4 t_3^2 \right. \nonumber\\
&&+  \frac{\Delta^2}{2}  \left(1-t_3^2\right) G^{\mu\nu} \mathcal J_\mu \mathcal J_\nu +
\frac{\Delta^2}{2} G^{\mu\nu} \nabla_\mu \mathbf t \! \cdot \! \nabla_\mu \mathbf t \nonumber\\
&&\left. + \frac{\Delta^2}{4} g_{\alpha \beta} G^{\mu\nu} G^{\lambda \eta} \left( {{\mathcal D_\mu}^\sigma}_\lambda 
\ce_\sigma \right)^\alpha \left({{\mathcal D_\nu}^\kappa}_\eta
\ce_\kappa \right)^\beta \right\}\nonumber\\
\label{covarYM}
\end{eqnarray} 
where
\begin{eqnarray*}
{\mathcal F}_{\mu\nu} & = & \partial_\mu \mathcal J_\nu - \partial_\nu \mathcal J_\mu 
+  \Delta \, t_3 \, \mathcal H_{\mu\nu}\nonumber\\
\mathcal H_{\mu\nu} 
&= & \frac{i}{2} \left( \ce_{\mu} \bar\ce_{\nu} -  \ce_{\nu} \bar\ce_{\mu} \right).
\end{eqnarray*}
Note that, the Lagrangian (\ref{covarYM}) has a manifestly covariant form, but by construction it presumes that the  underlying manifold is locally conformally flat. 
Thus,  the two manifolds  $\mathcal M$ and  $\mathcal N$ alluded to at the beginning of the present section have to be chosen  so that the product $\mathcal M \times \mathcal N$ is locally conformally flat.

\section{Concluding remark}

In this paper a reformulation of the Frenet equation in order to describe three dimensional strings in terms of spinors is presented. In addition, it is shown that an extension of   the spinor Frenet equation to include  
the time evolution of strings leads to a Maurer-Cartan structure, which is  related 
to the Lax pair representation of  two dimensional integrable models. 
Since the  time evolution of  a string determines a  Riemann surface in $\mathbb R^3$, it has been shown that there is  a direct connection between  the spinor representation of Frenet equation and  of the Gau\ss-Godazzi equations.

Finally,   the relations between strings, surfaces and integrable models 
in terms of decomposed representations of both two dimensional and four dimensional SU(2)  Yang-Mills theories has been studied in detail.
 This study indicates  that the four dimensional decomposed Yang-Mills  in combination with Cartan geometry, provides a unifying framework to describe string dynamics,  the structure of  Riemann surfaces, and two dimensional integrable models.

\section{Acknowledgements:}
T.I. thanks Tours University, Uppsala University, Beijing Institute of Technology and 
Shanghai University for hospitality. Y.J. thanks
Tours University for hospitality. A.J.N. thanks Shanghai University for hospitality.

T.I. acknowledges  support from FP7, Marie Curie Actions, People, International Research Staff Exchange Scheme (IRSES-606096) and  from The Hellenic Ministry of Education: Education and Lifelong Learning Affairs, and European Social Fund: NSRF 2007-2013, Aristeia (Excellence) II (TS-3647).
YJ acknowledges support from the National Natural Science Foundation of China (11275119),  the Ph.D. Programs Foundation of Ministry of Education of China (20123108110004), and Sino-French Cai Yuanpei Exchange Program.
A.J.N.  acknowledges  support from CNRS PEPS grant, Region Centre 
Rech\-erche d$^{\prime}$Initiative Academique grant, Sino-French Cai Yuanpei Exchange Program (Partenariat Hubert Curien), Vetenskapsr\aa det, Carl Trygger's Stiftelse f\"or vetenskaplig forskning, and  Qian Ren Grant at BIT.

\vfill\eject


\begin{thebibliography}{99}
\bibitem{foot} We note that 
if instead of $\mathbb R^3$ the ambient space is a three dimensional curved manifold, 
the condition (\ref{Fts}) must be adjusted so that it accounts for the nontrivial curvature of the 
ambient space \cite{oma3d}.
\end{thebibliography}
\end{document}